\documentclass[12pt,draftclsnofoot,onecolumn]{IEEEtran}
\IEEEoverridecommandlockouts
\usepackage[T1]{fontenc}
\newif\ifhavebib

\usepackage{blindtext}
\usepackage{graphicx}
\usepackage{graphicx,subfigure}
\usepackage{array}
\usepackage{url}
\usepackage{algorithmic}
\usepackage[cmex10]{amsmath}
\usepackage{ifpdf}
\usepackage{amssymb,amsmath}
\usepackage{empheq}
\usepackage{stfloats}   
\usepackage{textcomp}
\usepackage{cite}
\usepackage{multirow}
\usepackage{diagbox}
\let\oldFootnote\footnote
\newcommand\nextToken\relax

\renewcommand\footnote[1]{%
	\oldFootnote{#1}\futurelet\nextToken\isFootnote}

\newcommand\isFootnote{%
	\ifx\footnote\nextToken\textsuperscript{,}\fi}

\usepackage[ruled,vlined,lined,ruled,boxed]{algorithm2e}
\usepackage{fixltx2e}
\usepackage[dvipsnames,usenames]{color}
\usepackage[normalem]{ulem}
\usepackage{amsthm}

\usepackage{graphicx}
\usepackage{ifpdf}
\usepackage{cite}
\usepackage{enumerate}
\usepackage{enumitem}

\usepackage{array}
\usepackage{mdwmath}
\usepackage{mdwtab}
\usepackage{eqparbox}
\usepackage{bm}

\usepackage{setspace}
\usepackage{booktabs}
\usepackage[subnum]{cases}
\usepackage{rotating}

\usepackage{listings} 
\definecolor{Red}{rgb}{1,0,0}
\definecolor{Blue}{rgb}{0,0,1}
\definecolor{Olive}{rgb}{0.41,0.55,0.13}
\definecolor{Green}{rgb}{0,1,0}
\definecolor{MGreen}{rgb}{0,0.8,0}
\definecolor{DGreen}{rgb}{0,0.55,0}
\definecolor{Yellow}{rgb}{1,1,0}
\definecolor{Cyan}{rgb}{0,1,1}
\definecolor{Magenta}{rgb}{1,0,1}
\definecolor{Orange}{rgb}{1,.5,0}
\definecolor{Violet}{rgb}{.5,0,.5}
\definecolor{Purple}{rgb}{.75,0,.25}
\definecolor{Brown}{rgb}{.75,.5,.25}
\definecolor{Grey}{rgb}{.5,.5,.5}

\def\blue{\color{Blue}}

\newcommand{\boxhead}[5]{
	\pagestyle{myheadings}
	\thispagestyle{plain}
	\setcounter{page}{1}
	\noindent
	\begin{center}
		\framebox{
			\vbox{\vspace{2mm}
				\hbox to 6.28in { {\bf #1 \hfill} }
				\vspace{6mm}
				\hbox to 6.28in { {\Large \hfill \bf #2  \hfill} }
				\vspace{6mm}
				\hbox to 6.28in { {\it #3 #4 \hfill  #5} }
				\vspace{2mm}}
		}
	\end{center}
	\markboth{#5 -- #2}{#5 -- #2}
	\vspace*{4mm}
}

\usepackage[colorlinks=true, urlcolor=black,  linkcolor=black, citecolor=black]{hyperref}
\theoremstyle{definition}
 
\newtheorem{proposition}{Proposition} 
\newtheorem{corollary}{Corollary}

\theoremstyle{remark}
\newtheorem{remark}{Remark}

\theoremstyle{definition}

\DeclarePairedDelimiterX{\infdivx}[2]{(}{)}{%
	#1\;\delimsize\|\;#2%
}

\DeclarePairedDelimiter{\norm}{\lVert}{\rVert}
\DeclarePairedDelimiter{\abs}{\lvert}{\rvert}
\DeclareMathOperator*{\argmax}{argmax}
\DeclareMathOperator*{\argmin}{argmin}
\newcommand{\av}{{\bf a}}

\newcommand{\Xv}{{\bf X}}
\newcommand{\Yv}{{\bf Y}}

\newcommand{\Fv}{{\bf F}}
\newcommand{\Vv}{{\bf V}}
\newcommand{\Rv}{{\bf R}}

\newcommand{\Mv}{{\bf M}}
\newcommand{\Av}{{\bf A}}
\newcommand{\rv}{{\bf r}}

\newcommand{\Pv}{{\bf P}}
\newcommand{\Hv}{{\bf H}}
\newcommand{\Gv}{{\bf G}}
\newcommand{\Sv}{{\bf S}}
\newcommand{\Nv}{{\bf N}}

\newcommand{\Iv}{{\bf I}}
\newcommand{\fv}{{\bf f}}

\newcommand{\xv}{{\bf x}}

\newcommand{\qv}{{\bf q}}

\newcommand{\uv}{{\bf u}}
\newcommand{\vv}{{\bf v}}
\newcommand{\hv}{{\bf h}}
\newcommand{\sv}{{\bf s}}
\newcommand{\nv}{{\bf n}}
\newcommand{\deltav}{\boldsymbol \delta}

\newcommand{\gammav}{\boldsymbol \gamma}

\newcommand{\thetav}{\boldsymbol \theta}
\newcommand{\varthetav}{\boldsymbol \vartheta}
\newcommand{\muv}{\boldsymbol \mu}
\newcommand{\Sigmav}{\boldsymbol \Sigma}

\DeclareMathOperator\E{E}

\def\E{\mathbb{E}}

\def\de \mathrm{d}

\newcommand{\Norm}{\mathcal{N}}
\newcommand{\CN}{\mathcal{CN}}

\newcommand\eg{e.g.,\xspace}
\newcommand\ie{i.e.,\xspace}
\def\textiid{i.i.d.\@\xspace}
\newcommand\iid{\ifmmode\text{ i.i.d. } \else \textiid \fi}

\newcommand{\Complex}{\mathbb{C}}
\newcommand{\Real}{\mathbb{R}}
\newcommand{\half}{\frac{1}{2}}

\newcommand{\beqs}{\begin{equation*}}
\newcommand{\eeqs}{\end{equation*}}
\newcommand{\beq}{\begin{equation}}
\newcommand{\eeq}{\end{equation}}
\begin{document}
\setitemize{listparindent=\parindent,partopsep=0pt,topsep=-0.25ex}
\setenumerate{fullwidth,itemindent=\parindent,listparindent=\parindent,itemsep=0ex,partopsep=0pt,parsep=0ex}
\havebibtrue
\title{Statistical Beamforming for FDD Downlink Massive MIMO via Spatial Information Extraction and Beam Selection}
\author{Hang~Liu, Xiaojun~Yuan,~\IEEEmembership{Senior Member,~IEEE,}
	and~Ying~Jun (Angela)~Zhang,~\IEEEmembership{Fellow,~IEEE}
	\thanks{This work was presented in part at the IEEE Global Communications Conference (GLOBECOM), Waikoloa, Hawaii, USA, Dec. 2019 \cite{Globecompaper}.

		H. Liu and Y. J. Zhang are with the Department of Information Engineering, The Chinese University of Hong Kong, Shatin, New Territories, Hong Kong (e-mail: lh117@ie.cuhk.edu.hk; yjzhang@ie.cuhk.edu.hk). X. Yuan is with the Center for Intelligent Networking and Communications, the University of Electronic Science and Technology of China, Chengdu, China (e-mail: xjyuan@uestc.edu.cn).}
}
\maketitle
\begin{abstract}
In this paper, we study the beamforming design problem in frequency-division duplexing (FDD) downlink massive MIMO  systems, where instantaneous channel state information (CSI) is assumed to be unavailable at the base station (BS). We propose to extract the information of the angle-of-departures (AoDs) and the corresponding large-scale fading coefficients (a.k.a. spatial information) of the downlink channel from the uplink channel estimation procedure, based on which a novel downlink beamforming design is presented. By separating the subpaths for different users based on the spatial information and the hidden sparsity of the physical channel, we construct near-orthogonal virtual channels in the beamforming design. Furthermore, we derive a sum-rate expression and its approximations for the proposed system. Based on these closed-form rate expressions, we develop two low-complexity beam selection schemes and carry out asymptotic analysis to provide valuable insights on the system design. Numerical results demonstrate a significant performance improvement of our proposed algorithm over the state-of-the-art beamforming approach. 
\end{abstract}
\begin{IEEEkeywords}
	Massive MIMO, FDD, statistical beamforming, spatial reciprocity.
\end{IEEEkeywords}
\section{Introduction}\label{sec_intro}
Massive multiuser multiple-input multiple-output (MIMO) that employs a large-scale antenna array at the base station (BS) to serve multiple users  has been extensively studied for its remarkable enhancement in system throughput \cite{MIMO_over1, MIMO_over3, MIMO_over4, MIMO_over5}.  
To achieve the high array gain of a massive MIMO system, the channel state information (CSI) at the BS is crucial for uplink signal detection and downlink precoding. Thanks to the powerful signal processing capability of the BS, the uplink channel coefficients can be learned accurately even with very short pilot sequences, provided that channel structural information, such as the hidden channel sparsity, can be efficiently exploited \cite{MIMO_zhangjianwen,MIMO_leichen, Superreso_arxiv,MIMO_yanwenjing}.
 Meanwhile, the downlink CSI for time division duplexing (TDD) systems is usually obtained from a direct reuse of the uplink CSI under the assumption of the uplink/downlink channel reciprocity \cite{TDD_reuse}. On the contrary, the downlink CSI acquisition in frequency division duplexing (FDD) systems is much more challenging, since the channel reciprocity is not applicable therein \cite{MIMO_DTse}.  In order to learn the downlink CSI, the BS needs to broadcast pilot signals with the length proportional to the number of transmit antennas. After receiving the pilots, every user estimates the channel coefficients and feeds back the quantized CSI to the BS, which leads to additional channel errors and feedback overheads.

Most existing cellular systems operate in FDD, thanks to its advantages for systems with symmetric uplink/downlink traffic and/or low-delay demands \cite{FDD_VS_TDD}. Therefore, it is of utmost importance for the FDD systems to efficiently acquire the downlink CSI at the transmitter (CSIT). Many studies have been conducted to relieve the overwhelming training and feedback burden in the conventional training-based framework. For example, Ref. \cite{TSBF1} proposed a two-stage beamforming (TSBF) algorithm consisting of inner and outer precoding processes. First, an outer precoder (a.k.a. prebeamforming matrix) determined by the channel statistics is employed to partition users into groups, where pilots can be reused in different groups. Then, an inner precoder is designed on top of the product of the channel and the outer precoder after downlink training. Another line of research, namely statistical beamforming (SBF), designs the downlink system merely based on the channel covariance matrix (CCM) without resorting to downlink training. SBF was initially considered with restriction to a simple two-user scenario, by optimizing the ergodic sum-rate \cite{SBF_2.5} or the signal-to-leakage-and-noise ratio (SLNR) \cite{SBF_1.5}. These results are later extended to the massive MIMO systems in \cite{SBF_2} and \cite{SBF1}. Moreover, Ref. \cite{SBF3_2} employs a discrete Fourier transform (DFT) basis to simplify the SLNR maximization in \cite{SBF_1.5}.  {The CCM-based algorithms usually acquire the uplink CCM information via channel averaging and construct the downlink CCM from the uplink CCM. 
{\blue However, the transformation from the uplink CCM to the downlink CCM is non-trivial, for that the uplink and downlink communications occur over different frequency bands in the FDD massive MIMO systems \cite{MIMO_GCaire2}.}

Experimental studies in \cite{Reciprocity_exp} have evidenced the congruence of the directional properties of the uplink and downlink channels. {\blue Motivated by this observation, the authors in \cite{MIMO_Xie,MIMO_GCaire} proposed to extract the spatial information and use the extracted information to reconstruct the downlink CCM, under the assumption of spatial reciprocity (\ie angle reciprocity and power angular spectrum (PAS) reciprocity) between the uplink and downlink channels.\footnote{{In this paper, ``channel reciprocity” refers to the phenomenon that the uplink and downlink channels are identical. By contrast, ``spatial reciprocity” refers to the congruence of the angle parameters and the PASs in the uplink and downlink channels.}} Besides, Ref. \cite{MIMO_DOA1,MIMO_Ding} proposed to jointly estimate the uplink and downlink channels, where the uplink training is used to help improving the downlink channel estimation accuracy under the assumption of angle reciprocity. } 

In this paper, we are interested in the downlink beamforming design for FDD massive MIMO without downlink training. We propose to acquire the spatial information from a blind uplink channel-and-signal estimation process \cite{Superreso_arxiv} and formulate the corresponding problem as an SBF design task.
We develop a novel beamforming algorithm that assigns a certain number of angular beams to each supported user in the precoding. Furthermore, a sum-rate expression and its approximations are derived for the proposed algorithm. Based on this, we develop two beam selection algorithms and analytically characterize the sum-rate performance. As a consequence, the overall framework hinges on the spatial information but not on the instantaneous CSIT. 
Moreover, the proposed design acquires the spatial information from a tunable sampling basis, and thus can alleviate the energy leakage problem in the DFT-based methods. The main contributions of this paper are summarized as follows.
\begin{itemize}
	\item We study the downlink beamforming design problem for FDD massive MIMO with no instantaneous CSIT. By exploiting the spatial reciprocity, we develop an  angle-of-departure (AoD) and PAS extraction framework based on the estimates of the uplink counterparts. It is worth mentioning that this framework, together with the proposed beamforming algorithm, eliminates the training and feedback overheads of the CSIT acquisition required in the conventional beamforming approach.
	\item We propose a novel beamforming design to assign angular beams to different users. Moreover, we develop an associated beam-time block coding scheme to tackle the problem of the potential phase destruction, which is caused by the unknown channel phase coefficients of multiple beams assigned to a single user.
	\item We develop two beam selection schemes based on forward stepwise searching and Gibbs sampling. By maximizing a tight approximation of the sum-rate, the low-complexity beam selection schemes significantly enhance the performance of the proposed beamforming system.
	\item We derive an exact sum-rate expression for the proposed system, and provide asymptotic analysis to shed light on the system design. The advantages of the proposed framework are verified by extensive numerical comparisons with the existing training-based beamforming and the CCM-based SBF.
\end{itemize}

The remainder of this paper is organized as follows. In Section \ref{sec_model}, we describe the multiuser massive MIMO system model. In Section \ref{sec3}, we first investigate the inherent uplink/downlink spatial reciprocity of the physical channel. Based on this, the spatial information extraction framework is discussed and compared with the existing approaches. In Section \ref{sec_DSBF}, we introduce the proposed beamforming algorithm and derive an exact sum-rate expression. The rate expression is utilized to design the associated beam selection schemes in Section \ref{sec3.3}. In Section \ref{sec_asymoptotic}, we present the asymptotic analysis and other analytical results on the sum-rate performance. Furthermore, Section \ref{sec_simulation} gives numerical results of the proposed methods. Finally, the paper concludes in Section \ref{conclusions}.

\emph{Notation}: Throughout, we use $\Complex$ and $\Real$ to denote the real and complex number sets, respectively. Regular small letters, bold small letters, and bold capital letters are used to denote scalars, vectors, and matrices, respectively. We use $x_{ij}$ to denote the $(i,j)$-th entry of matrix $\Xv$. We use $(\cdot)^\star$, $(\cdot)^T$, $(\cdot)^H$, $(\cdot)^{-1}$ to denote the conjugate, the transpose, the conjugate transpose, and the inverse, respectively. We use $\Norm(\cdot;\muv,\Sigmav)$ and $\CN(\cdot;\muv,\Sigmav)$ to denote the real normal and the circularly-symmetric normal distributions with mean $\muv$ and covariance $\Sigmav$, respectively. 
We use $\E[\cdot]$ to denote the expectation operator, $\abs{\cdot}$ to denote the cardinality of a set or a multiset, $\norm{\cdot}_p$ to denote the $\ell_p$ norm, $\Iv_N$ to denote the $N\times N$ identity matrix, and $diag(\xv)$ to denote the diagonal matrix with diagonal entries specified by $\xv$. 
Finally, we define $[n] \triangleq \{1,2,3,\cdots,n \}$ for some positive integer $n$.
\section{System Model}
\label{sec_model}
We consider a downlink massive MIMO system in a single cell. An $N$-antenna BS serves $K$ single-antenna users, where $N \!\gg\! K \!\gg\! 1$. Let $\xv=[x_1,x_2,\cdots,x_K]^T\in \Complex^{ K\times 1}$ denote the vector of signals that the BS transmits to the users, where $x_k$ is the signal desired by the $k$-th user with $\E[\abs{x_k}^2]=1$. Before transmission, the signals are linearly precoded by a beamforming matrix $\Vv=[\vv_1,\vv_2,\dots,\vv_K] \in \Complex^{N\times K}$, where $\norm{\vv_k}_2^2=1, \forall k$ without loss of generality. The received signal at the $k$-th user is given by
\begin{align}
\label{yk}
y_k&=\hv_k^H\Vv diag(\gammav)^{\frac{1}{2}} \xv+n_k, \nonumber\\
&=\sqrt{\gamma_k} \hv_k^H \vv_k x_k+\sum_{j\neq k }\sqrt{\gamma_j}\hv_k^H\vv_j x_j+n_k,
\end{align}
where $\hv_k \in \Complex^{N\times 1}$ is the complex-valued channel coefficient vector between the BS and the $k$-th user; $n_k$ is the additive white Gaussian noise (AWGN) following an independent and identically distributed (\iid) complex Gaussian distribution $\CN(0,1)$; and $\gammav\triangleq [\gamma_1,\gamma_2,\cdots,\gamma_K]^T$ is the power allocation vector. We assume equal power allocation for all users, namely, $\gamma_1=\gamma_2=\cdots=\gamma_K=P/K\triangleq\gamma$ with total transmission power $P$.

The downlink channel $\hv_k$ in \eqref{yk} can be modeled as \cite{Channel_formular}
\begin{align}
\label{downlinkchannel}
\hv_k&= \sum_{p=1}^{P(k)} \av(\theta_{k,p},f) \varsigma_{k,p}s_{k,p},
\end{align}
where $f$ is the downlink carrier frequency; $P(k)$ is the total number of resolvable propagation paths of the channel between the BS and the $k$-th user; $\theta_{k,p}$ is the corresponding AoD; $\varsigma_{k,p}$ is the corresponding large-scale fading coefficient, \ie $\{\varsigma^2_{k,p}: p \in [P(k)]\}$ represents the PAS of the $k$-th user \cite{PAS}; $s_{k,p}\sim \CN(0, 1)$ is the corresponding small-scale fading coefficient;  and $\av(\theta,f)$ is the steering vector for a signal impinging upon the antenna array at AoD $\theta$ and frequency $f$. {\blue We assume a uniform linear array (ULA) with isotropic antenna elements at the BS.} The corresponding steering vector is given by
\begin{equation}
\label{ULA}
\av(\theta,f)\!=\!\frac{1}{\sqrt{N}}\!\left[1,e^{-j\frac{2\pi f}{c_0}d\sin(\theta)},\cdots,e^{-j\frac{2\pi f}{c_0}d(N-1)\sin(\theta)}\right]^T\!,
\end{equation}
where $c_0$ denotes the speed of light, and $d$ denotes the distance between any two adjacent antennas.

Define the collection of all AoDs as 
\begin{align}
\thetav\triangleq \{\theta_{k,p}:p\in P(k),k \in [K]\}.
\end{align}
With respect to an angle grid $\varthetav=\{\vartheta_l: l \in [L]\}$ {\blue with length $L$ ($\leq N$)}\footnote{\blue We assume that $N$ is sufficiently large, and that the grid $\varthetav$ with length no greater than $N$ samples the sparse channel vectors in the angular domain.} that covers the AoD range $[-90^\circ,90^\circ]$, the array response matrix is denoted by $\Av(\varthetav,f)=[\av(\vartheta_1,f),\av(\vartheta_2,f),\dots,\av(\vartheta_{L},f)]$. 
We note that if $\thetav$ is well covered by $\varthetav$, the channel in \eqref{downlinkchannel} can be represented under the basis $\Av(\varthetav,f)$ as
\begin{align}
\label{downlinkgrid}
\hv_k&=\Av(\varthetav,f)\Sigmav_k\sv_k,
\end{align}
where $\sv_k \sim \CN({\bf 0},\Iv_{L})$; and  $\Sigmav_k$ is a diagonal matrix with the $l$-th diagonal entry given by
\begin{align}\label{temp50}{\sigma_{k,l}\!=\!}
\begin{cases}
\varsigma_{k,p}, &\text{if } \vartheta_l\!= \!\theta_{k,p} \text{ for some } \theta_{k,p}\! \in\! \thetav, \\
0, &\text{otherwise.}
\end{cases}
\end{align}
Define the available beam set for the $k$-th user as 
\begin{align}\label{sk}
\mathcal{S}_k=\{l \in [L]: \sigma_{k,l}\neq 0 \}.
\end{align}
With this definition, we rewrite \eqref{downlinkgrid} as 
\begin{align}
\label{downlink}
\hv_k&=\sum_{l\in \mathcal{S}_k}\av(\vartheta_l,f) \sigma_{k,l}s_{k,l}.
\end{align}
Due to a limited number of scatterers in the physical environment, the propagation channel exhibits a sparse structure in the angular domain\cite{Channel_Geometry,Channel_3DmmWave}, \ie $\abs{\mathcal{S}_k}\ll L$. 
It is worth noting that there may exist some mismatch between $\thetav$ and $\varthetav$ in practice, \ie some AoDs are not on the grid $\varthetav$. Following \cite{MIMO_DOA1}, we model the mismatch by $\deltav=\{\delta_l: l \in [L]\}$. Specifically, if $\theta_{l^\prime} \notin \varthetav$ and $\vartheta_l$ is the nearest grid point to $\theta_{l^\prime}$, we have 
\begin{align}
\label{offgridmismatch}
\vartheta_l=\theta_{l^\prime}-\delta_l.
\end{align}
%
{\blue When $\deltav\neq \bf 0$, the channel representation in \eqref{downlink} is inexact and the channel coefficient vector in the angular domain is not exactly sparse. In other words, the energy of the beams in $\mathcal{S}_k$ leaks to the nearby beams due to the angle mismatch. Studies on the angle mismatch phenomenon is postponed to Section \ref{sec_simulation2}, after we introduce the proposed beamforming scheme.
}

Under the block-fading assumption, the ergodic achievable rate of user $k$ for a given $\Vv$ can be expressed as
\begin{align}
\label{temp64}
R_k&=\E\left[\log_2 \left( 1+\text{SINR}_k\right) \right],
\end{align}
where the signal-to-interference-plus-noise ratio (SINR) of user $k$ is given by
\begin{align}
\label{SINR}
\text{SINR}_k&=\frac{\gamma \abs{ \hv_k^H \vv_k}^2}{1+\gamma \sum_{j\neq k}\abs{ \hv_k^H \vv_j}^2}.
\end{align}
The ergodic sum-rate is then given by
\begin{align}
\label{temp57}
R_{\text{sum}}=\sum_{k=1}^K R_k.
\end{align}

The maximization of $R_{\text{sum}}$ in \eqref{temp57} over the beamforming matrix $\Vv$ generally requires the knowledge of CSIT. As discussed in the Introduction, the training-based CSIT acquisition method may become infeasible due to the unaffordable training overhead and the inevitable channel errors in the CSI quantization and feedback processes. Although compressed-sensing (CS) based techniques can be employed to reduce the training overhead \cite{Channel_virtual_representation,MIMO_pilot6}, the downlink training process may still be resource-consuming, for that the required pilot length is still proportional to the number of transmit antennas to ensure a diminishing estimation error at the high signal-to-noise ratio (SNR) regime. Motivated by this, we propose to extract the spatial information from the uplink system for the downlink beamforming design.

\section{Uplink/Downlink Spatial Reciprocity and Spatial Information Extraction}
\label{sec3}
In this section, we develop a method for spatial information (\ie $\{\theta_{k,p}\}$ and $\{\varsigma_{k,p}\}$) extraction by exploiting the uplink/downlink spatial reciprocity and our prior work on the uplink channel estimation \cite{Superreso_arxiv}.
\subsection{Spatial Reciprocity}
Similar to \eqref{downlinkgrid}, the uplink channel can be expressed as 
\begin{align}
\label{uplinkchannel}
\hv_k^{\text{ul}}&=\sum_{l=1}^{L^{\text{ul}}} \av(\vartheta_l^{\text{ul}},f^{\text{ul}}) \sigma_{k,l}^{\text{ul}}s_{k,l}^{\text{ul}},
\end{align}
where $f^{\text{ul}}$ is the uplink carrier frequency; $\varthetav^{\text{ul}}=\{\vartheta_l^{\text{ul}}\}$ is the grid sampling the uplink angle-of-arrivals (AoAs) with length $L^{\text{ul}}$; $\sigma_{k,l}^{\text{ul}}$ and $s_{k,l}^{\text{ul}}$ are the corresponding large-scale and small-scale fading coefficients under the array response $\Av(\varthetav^{\text{ul}},f^{\text{ul}})=[\av(\vartheta_1^{\text{ul}},f^{\text{ul}}),\dots,\av(\vartheta_{L}^{\text{ul}},f^{\text{ul}})]$.

Although the channel reciprocity is not applicable to the FDD systems \cite{MIMO_DTse}, the experiments in \cite{Reciprocity_exp} demonstrated the congruence of the dominant AoA/DoA and the high correlation between the uplink and downlink PASs when the frequency duplex distance is small. Following \cite{MIMO_Ding,MIMO_DOA1,MIMO_Xie,MIMO_GCaire}, we assume that the AoA/AoD reciprocity and the PAS reciprocity hold between the uplink and downlink channels, \ie $\thetav$ and $\{\varsigma_{k,p}\}$ are identical to their uplink counterparts.
\subsection{Spatial Information Extraction}
With spatial reciprocity, we now discuss how to extract the downlink spatial information $\thetav$ and  $\{\varsigma_{k,p}\}$ (or equivalently, $\varthetav$ and $\{\sigma_{k,l}\}$) from the uplink channel.
In \cite{Superreso_arxiv}, we developed a blind channel-and-signal estimation algorithm for the uplink massive MIMO systems, where the estimation of the channel parameters, including $\varthetav^{\text{ul}}$, are involved in the algorithm. Specifically, denote by $\Xv^{\text{ul}} \in \Complex^{K\times T}$ the collection of the uplink signals. The received signal matrix $\Yv^{\text{ul}}$ is given by
\begin{align}
\Yv^{\text{ul}}&=\Hv^{\text{ul}}\Xv^{\text{ul}}+\Nv^{\text{ul}}=\Av(\varthetav^{\text{ul}},f^{\text{ul}}){\Sv}^{\text{ul}}\Xv^{\text{ul}}+\Nv^{\text{ul}},
\end{align}
where $\Hv^{\text{ul}}=[\hv_1^{\text{ul}},\dots,\hv_K^{\text{ul}}]$ with each $\hv_k^{\text{ul}}$ defined in \eqref{uplinkchannel}; $\Nv^{\text{ul}}$ is an AWGN matrix; and $ \Sv^{\text{ul}}$ is the channel coefficient matrix in the angular domain. The uplink estimation algorithm infers the posterior of $\Xv^{\text{ul}}$ and $\Sv^{\text{ul}}$ given $\Yv^{\text{ul}}$ and tunes the model parameters in an alternating fashion. The details can be found in 
\cite[Section III]{Superreso_arxiv}. 

{\blue Apart from $\varthetav^{\text{ul}}$, we also require the uplink large-scale fading components $(\sigma^{\text{ul}}_{k,l} )^2=\E[\abs{s_{k,l}^{\text{ul}}}^2]$. Specifically,  $(\sigma_{k,l}^{\text{ul}})^2$ can be computed as the sample average of $\{\abs{s_{k,l}^{\text{ul}}}^2\}$ by channel realizations of multiple coherence blocks. Alternatively, if a single coherence block is considered and hence only one instantaneous channel realization is available, $(\sigma_{k,l}^{\text{ul}})^2$ can be approximated by a single realization $\abs{s_{k,l}^{\text{ul}}}^2$. Note that this one-shot approximation is also employed in \cite[Section III-B]{MIMO_Xie}.}

Once the values of $\varthetav^{\text{ul}}$ and $\{\sigma_{k,l}^{\text{ul}}\}$ are obtained, we set 
\begin{align}\label{temp01}
\varthetav=\varthetav^{\text{ul}}, \sigma_{k,l}=\sigma^\text{ul}_{k,l}.
\end{align} 
\remark{\blue If $L^{\text{ul}}> L$ for some specific $L$, we additionally truncate the chosen $\varthetav$ and $\{\sigma_{k,l}\}$ to make sure they have $L$ elements. Specifically, we sequentially discard $\vartheta_{l^\prime}$ from $\varthetav$ and $\sigma_{k,l^\prime}$ in $\{\sigma_{k,l}\}$, where $l^\prime=\argmin_l \sum_k \sigma_{k,l}^2$. We repeat the truncation procedure until $\varthetav$ and $\{\sigma_{k,l}:\forall l\}$ exactly have $L$ elements.
}
\remark{Compared with the existing work \cite{MIMO_Ding,MIMO_DOA1,MIMO_Xie,MIMO_GCaire} that utilizes the spatial reciprocity to enhance the uplink and/or downlink training, our proposed framework extracts the spatial information directly from the uplink blind signal estimation process, which significantly reduces the training overhead in the uplink.\footnote{Strictly speaking, the uplink blind signal estimation in \cite{Superreso_arxiv} still requires a certain amount of training resources, since a short pilot sequence is inserted into each user packet to eliminate the phase and permutation ambiguities inherent in sparse matrix factorization.} Moreover, since we aim to design the beamforming based on $\varthetav$ and $\{\sigma_{k,l}\}$ only, the closed-loop downlink training is no longer necessary. More detailed comparisons are presented as follows.
\begin{itemize}
	\item In \cite{MIMO_Xie,MIMO_GCaire}, the spatial information is utilized to construct the downlink CCMs. {\blue Specifically,  Ref. \cite{MIMO_Xie} extracts the spatial information from the estimates of the uplink channels and uses the information to infer the downlink CCMs with the assumption of the spatial reciprocity. Differently, Ref. \cite{MIMO_GCaire} first computes the uplink sample covariance matrices with respect to (w.r.t.) the uplink channel estimates. The uplink CCM for user $k$, denoted by $\Rv_k(f^{\text{ul}})$, is estimated by projecting the corresponding sample covariance to the Toeplitz, positive semidefinate cone. Then, the downlink CCM for user $k$, denoted by $\Rv_k(f)$, is estimated from the uplink CCM, with the assumption of the spatial reciprocity.} 
	\item In \cite{MIMO_Ding,MIMO_DOA1}, uplink channel estimates are used to enhance the downlink channel training, so as to reduce the training overhead.
	However, as discussed in Section \ref{sec_model}, the downlink training may still degrade the overall performance. 
	Moreover, Ref. \cite{MIMO_Ding} utilizes the angle reciprocity by assuming  $ \av(\theta_{k,p},f)= \av(\theta_{k,p}^{\text{ul}},f^{\text{ul}})$. This assumption is questionable since usually we have $f \neq f^{\text{ul}}$.
	\item We emphasize that the SBF design proposed in this paper merely requires the spatial information. This implies that the existing mechanisms \cite{MIMO_Ding,MIMO_DOA1,MIMO_Xie,MIMO_GCaire} that can extract the DoAs $\thetav$ (or the grid $\varthetav$) and the large-scale fading coefficients $\{\varsigma_{k,p}\}$ (or $\{\sigma_{k,l}\}$) are all compatible with our proposed beamforming design. Here, the proposed approach based on \cite{Superreso_arxiv} is preferable since it does not involve additional channel training. 
\end{itemize}
}
\section{Beam-Selection-Based Statistical Beamforming}\label{sec_DSBF}
In this section, we consider the downlink beamforming design problem. We assume that the BS has no instantaneous CSIT and every user has perfect CSI of its own\footnote{In practice, CSI acquisition at the user side for the proposed system can be conducted very efficiently with a neglectable training overhead, since only the knowledge of a small number of effective channel coefficients is required. The details will be explained in Section \ref{sec_BTBC}.}.With the knowledge of $\varthetav$ and $\{\sigma_{k,l}\}$, we aim to design the beamformer $\Vv$. For simplicity of exposition, we assume perfect spatial information in the sequel, \ie $\varthetav$ and $\{\sigma_{k,l}\}$ are accurate. 
Based on this, we propose a beam-selection-based statistical beamforming (BS-SBF) algorithm, where each beamforming vector occupies a number of angular beams. The robustness of the proposed design against the spatial information mismatch is verified in Section \ref{sec_simulation2}.
\subsection{Beamforming}\label{sec_bf}
Define 
\begin{align}
\label{temp65}
\Av^\ddagger \triangleq \Av(\varthetav)(\Av(\varthetav)^H\Av(\varthetav))^{-1},
\end{align}
where the downlink carrier frequency $f$ is omitted for the ease of notation. By definition we have $\Av(\varthetav)^H\Av^\ddagger =\Iv_{L}$.

In the proposed BS-SBF, each user selects $\Gamma>0$ distinct directions from the beam set $[L]$. Denote by $\mathcal{G}_k$ the selected directions. We have $\abs{\mathcal{G}_k}=\Gamma$ and $\mathcal{G}_i \bigcap \mathcal{G}_j=\emptyset, \forall i,j$. The selection of $\mathcal{G}_k$ can be found in Section \ref{sec3.3}. Given $\mathcal{G}_k$, the beamforming vector is expressed as
\begin{align}
\label{temp66}
\vv_k=\frac{1}{\sqrt{\Gamma}}\sum_{l\in \mathcal{G}_k} \av^\ddagger_l,
\end{align}
where $\av^\ddagger_l$ is the $l$-th column of $\Av^\ddagger$. From \eqref{downlink} and \eqref{temp66}, we have\footnote{\blue Without perfect angle information (\ie the angle mismatch vector $\deltav\neq \bf 0$), the beamforming vectors in \eqref{temp66} is inaccurate. As a result, the energy in the selected beams $\mathcal{G}_i \bigcap\mathcal{S}_k$ in \eqref{temp51} leaks to the nearby beams, which generally degrades the beamforming performance.}
\begin{align}
\label{temp51}
\gamma \abs{\hv_k^H\vv_i}^2=\frac{\gamma}{\Gamma}\Big \lvert \sum_{l\in \mathcal{G}_i\bigcap \mathcal{S}_k}\sigma_{k,l}s_{k,l}\Big \lvert^2.
\end{align}
When $\Gamma \!\geq\!2$, a potential phase mismatch issue arises in \eqref{temp51} due to the unknown $\{s_{k,l}\}$. Specifically, for the arbitrarily distributed phases of  $\{s_{k,l}\}$, the sum in \eqref{temp51} can be zero in the worst case when the phases of $\{s_{k,l}\}$ are destructive. To tackle this issue, we propose to combine BS-SBF with an angular-domain version of the space-time block coding (STBC), referred to as the beam-time block coding (BTBC). The details are described in the next subsection.
\subsection{BTBC}\label{sec_BTBC}
STBC is widely adopted to achieve a diversity gain without CSIT. Here, we extend the idea of STBC to the angular domain. For reasons presented later in Section \ref{sec_Gamma}, $\Gamma$ is usually a small number, \eg $1$ or $2$. Therefore, we take $\Gamma=2$ as an illustration of the BTBC, while higher order BTBCs can be derived with larger coding matrices \cite{STBC2} by following the same argument. 

Suppose that the selected beams for user $k$ are given by  $\mathcal{G}_k=\{l_1(k),l_2(k)\}$. Denote by $\qv=\Vv\xv=\sum_{k=1}^K \qv_k \in \Complex^{N\times1}$ the transmitted signal vector after precoding. We set $\qv_k(t)$ at time slot $t=1, 2$ as
\begin{align}
\qv_k(1)&=\frac{1}{\sqrt{2}}\av^\ddagger_{l_1(k)} x_k(1)+\frac{1}{\sqrt{2}} \av^\ddagger_{l_2(k)} x_k(2),\\
\qv_k(2)&=- \frac{1}{\sqrt{2}}\av^\ddagger_{l_1(k)}{x^\star_k(2)}+ \frac{1}{\sqrt{2}}\av^\ddagger_{l_2(k)} {x^\star_k(1)},
\end{align}
where $x_k(t)$ is the signal desired by the $k$-th user at time slot $t$.
The received signals at user $k$ are given by
\begin{align}
\label{temp61}
y_k(1)=&\sqrt{\frac{\gamma}{2}}\sigma_{k,l_1(k)}s_{k,l_1(k)}x_k(1)+\sqrt{\frac{\gamma}{2}}\sigma_{k,l_2(k)}s_{k,l_2(k)}x_k(2)\nonumber\\
&+\underbrace{\sqrt{\frac{\gamma}{2}}\! \sum_{\substack{j\neq k\\ l_1(j)\in\mathcal{S}_k }}\sigma_{k,l_1(j)}s_{k,l_1(j)}x_j(1)\!+\!\sqrt{\frac{\gamma}{2}} \!\sum_{\substack{j\neq k\\ l_2(j)\in\mathcal{S}_k }}\sigma_{k,l_2(j)}s_{k,l_2(j)}x_j(2)}_{\text{IUI}_k(1)}+n_k(1),\\
\label{temp62}
y_k(2)=&-\sqrt{\frac{\gamma}{2}}\sigma_{k,l_1(k)}s_{k,l_1(k)}{x^\star_k(2)}+\sqrt{\frac{\gamma}{2}}\sigma_{k,l_2(k)}s_{k,l_2(k)}{x^\star_k(1)}\nonumber\\
	&\underbrace{-\sqrt{\frac{\gamma}{2}}\! \sum_{\substack{j\neq k\\ l_1(j)\in\mathcal{S}_k }}\sigma_{k,l_1(j)}s_{k,l_1(j)}{x^\star_j(1)}\!+\!\sqrt{\frac{\gamma}{2}} \!\sum_{\substack{j\neq k\\ l_2(j)\in\mathcal{S}_k }}\sigma_{k,l_2(j)}s_{k,l_2(j)}{x^\star_j(2)}}_{\text{IUI}_k(2)}+n_k(2).
\end{align}
where the term contributing to the inter-user interference (IUI) at time slot $t$ is denoted by $\text{IUI}_k(t)$, for $t=1,2$. Collecting \eqref{temp61} and \eqref{temp62}, we obtain
\begin{align}
\begin{bmatrix}
y_k(1) \\
{y_k^\star(2)}
\end{bmatrix}=&\sqrt{\frac{\gamma}{2}}\underbrace{\begin{bmatrix}
	\sigma_{k,l_1(k)}s_{k,l_1(k)}&\sigma_{k,l_2(k)}s_{k,l_2(k)}\\
	\sigma_{k,l_2(k)}{s^\star_{k,l_2(k)}}&-\sigma_{k,l_1(k)}{s^\star_{k,l_1(k)}}
	\end{bmatrix}}_{\triangleq \Mv}
\begin{bmatrix}
x_k(1) \\
x_k(2)
\end{bmatrix}\nonumber\\
&+
\begin{bmatrix}
\text{IUI}_k(1)+n_k(1) \\
{\text{IUI}^\star_k(2)}+{n^\star_k(2)}
\end{bmatrix}.
\label{temp63}
\end{align}
In this case, the decoding at user $k$ can be conducted by multiplying $\sqrt{\frac{2}{\gamma}}\frac{\Mv^H}{\sigma^2_{k,l_1(k)}\abs{s_{k,l_1(k)}}^2+\sigma^2_{k,l_2(k)}\abs{s_{k,l_2(k)}}^2}$ at both sides of \eqref{temp63}. In general, user $k$ only needs to estimate $\Gamma$ effective channel coefficients along the selected beam directions, \ie $\{\sigma_{k,l}s_{k,l}\!:\!l\in \mathcal{G}_k\}$.  Since $\Gamma \!\ll\! N$, the training overhead is neglectable compared with that in the conventional training-based approach. 

\subsection{Sum-Rate Expression for the Proposed Scheme}
\label{sec_sumrate}
For $\Gamma=2$, $\text{SINR}_k$ in \eqref{SINR} under BS-SBF and BTBC is given by
\begin{align}
\label{SINR_us}
\text{SINR}_k&=\frac{\half \gamma\left( \sigma^2_{k,l_1(k)}\abs{s_{k,l_1(k)}}^2+\sigma^2_{k,l_2(k)}\abs{s_{k,l_2(k)}}^2\right) }
{1+\half \gamma\left( \sum_{\substack{j\neq k\\ l_1(j)\in\mathcal{S}_k }}\sigma^2_{k,l_1(j)}\abs{s_{k,l_1(j)}}^2+\sum_{\substack{j\neq k\\ l_2(j)\in\mathcal{S}_k }}\sigma^2_{k,l_2(j)}\abs{s_{k,l_2(j)}}^2\right) }.
\end{align}
For general $\Gamma>0$, $\text{SINR}_k$ is given by
\begin{align}
\label{SINR_Gamma}
\text{SINR}_k&=\frac{\frac{\gamma}{\Upsilon\Gamma}\sum_{l\in \mathcal{G}_k}\sigma^2_{k,l}\abs{s_{k,l}}^2}
{1+\frac{\gamma}{\Upsilon\Gamma}\sum_{j \neq k}\sum_{l\in \mathcal{G}_j\bigcap \mathcal{S}_k}\sigma^2_{k,l}\abs{s_{k,l}}^2 },
\end{align}
where $1/\Upsilon$ is the rate of the corresponding BTBC with $\Upsilon=1$ for $\Gamma=1,2$, and $\Upsilon>1$ for $\Gamma\geq 3$. 

Recall that by design we have  $\mathcal{G}_i \bigcap \mathcal{G}_j=\emptyset$ for  $\forall i,j\in [K]$. Define $\Omega_k \triangleq \mathcal{S}_k \bigcap \left( \bigcup_{k^\prime \in [K]} \mathcal{G}_{k^\prime}\right)$ to be the active beam set for user $k$, where $\mathcal{S}_k$ is defined in \eqref{sk}. 
We rearrange the terms in the denominator of \eqref{SINR_Gamma} and obtain
\begin{align}
\label{SINR_Gamma2}
\text{SINR}_k&=\frac{\frac{\gamma}{\Upsilon\Gamma}\sum_{l\in \mathcal{G}_k}\sigma^2_{k,l}\abs{s_{k,l}}^2}
{1+\frac{\gamma}{\Upsilon\Gamma}\sum_{l\in\Omega_k, l\notin \mathcal{G}_k}\sigma^2_{k,l}\abs{s_{k,l}}^2 }.
\end{align}	

With \eqref{SINR_Gamma2}, the exact expression of the achievable rate \eqref{temp64} is given in the following proposition. 
\begin{proposition}
{For each user $k$, $R_k=0$ when $\Omega_k = \emptyset$. Otherwise, suppose that the \emph{multiset}\footnote{Unlike a set, a multiset is a collection of elements, in which elements are allowed to
		repeat. For example, in multiset $\{a, a, b\}$, the element $a$ has multiplicity $2$, and $b$ has multiplicity $1$.} $\{\sigma_{k,l}:l\in \Omega_k\}$ has $J$ distinct values $\sigma_{k,1},\dots,\sigma_{k,J}$ with corresponding multiplicities $\rv=[r_{1},\dots,r_{J}]$, \ie $\sum_{j=1}^J r_j=\abs{\Omega_k}$. Similarly, suppose that the \emph{multiset} $\{\sigma_{k,l}: l \in \Omega_k, l \notin \mathcal{G}_k \}$ has $J^\prime$ distinct values $\sigma_{k,1},\dots,\sigma_{k,J^\prime}$ with corresponding multiplicities $\rv^\prime=[r^\prime_{1},\dots,r^\prime_{J^\prime}]$, \ie $\sum_{j^\prime=1}^{J^\prime} r_{j^\prime}=\abs{\Omega_k\setminus\mathcal{G}_k}$. The achievable rate $R_k$ is given by
	\begin{align}
\label{Rk_us_multi}
R_k&=\left(\prod_{j=1}^J \frac{1}{\sigma^{2r_j}_{k,j}} \right) \sum_{j=1}^J \sum_{l=1}^{r_j}\frac{(-1)^{2r_j-l-1}\sigma^{2(r_j-l+1)}_{k,j}}{\ln 2}f_{\rv,j,l}\exp\left(\frac{\Upsilon\Gamma}{\gamma\sigma^2_{k,j}} \right) \sum_{t=1}^{r_j-l+1}\!\!\! E_{t}\left( \frac{\Upsilon\Gamma}{\gamma\sigma^2_{k,j}}\right) \nonumber\\
&-\left(\prod_{j^\prime=1}^{J^\prime} \frac{1}{\sigma^{2r^\prime_{j^\prime}}_{k,j^\prime}} \right) \sum_{j^\prime=1}^{J^\prime} \sum_{l=1}^{r^\prime_{j^\prime}}\frac{(-1)^{2r^\prime_{j^\prime}-l-1}\sigma^{2(r^\prime_{j^\prime}-l+1)}_{k,{j^\prime}}}{\ln 2}f_{\rv^\prime,j^\prime,l}\exp\left(\frac{\Upsilon\Gamma}{\gamma\sigma^2_{k,{j^\prime}}} \right) \sum_{t=1}^{r^\prime_{j^\prime}-l+1}\!\!\! E_{t}\left( \frac{\Upsilon\Gamma}{\gamma\sigma^2_{k,{j^\prime}}}\right),
	\end{align}
where 
	\begin{align}
	f_{\rv,j,l}=\sum_{\substack{i_1,i_2,\dots,i_J: i_j=0\\ \sum_{\bar j=1}^J i_{\bar j}=l-1}}\prod_{\tau\in [J], \tau \neq j}\binom{i_\tau+r_\tau-1}{i_\tau}\left(\frac{1}{\sigma^2_{k,\tau}}-\frac{1}{\sigma^2_{k,j}} \right)^{-(r_{\tau}+i_\tau)},
		\end{align}
		and
		\begin{align}
		E_{t}(x)=\int_{1}^{\infty}\frac{e^{-x\zeta}}{\zeta^t}d\zeta.
		\end{align}
\label{pro1}}
\end{proposition}
\IEEEproof{See Appendix \ref{appa}.}
\corollary{For any user $k$, if the large-scale fading coefficients of the active beams $\{\sigma_{k,l}, l \in \Omega_k\}$ are distinct, $R_k$ can be simplified as \eqref{Rk_us} when $\Omega_k \neq \emptyset$.
	\begin{align}
	\label{Rk_us}
	R_k=& \frac{1}{\ln 2} \sum_{l \in \Omega_k}
	\left( \frac{1}{\prod_{\substack{j \neq l\\j \in \Omega_k}} \left( 1-\frac{\sigma^2_{k,j}}{\sigma^2_{k,l}}\right) } \right) \exp^{\frac{\Upsilon\Gamma}{\gamma \sigma^2_{k,l}}}E_1\left( \frac{\Upsilon\Gamma}{\gamma \sigma^2_{k,l}}\right) \nonumber\\
	&-\frac{1}{\ln 2} \sum_{{l^\prime} \in \Omega_k\setminus \mathcal{G}_k}
	\left( \frac{1}{\prod_{\substack{j \neq {l^\prime}\\j \in \Omega_k\setminus \mathcal{G}_k}} \left( 1-\frac{\sigma^2_{k,j}}{\sigma^2_{k,{l^\prime}}}\right) } \right) \exp^{\frac{\Upsilon\Gamma}{\gamma \sigma^2_{k,{l^\prime}}}}E_1\left( \frac{\Upsilon\Gamma}{\gamma \sigma^2_{k,{l^\prime}}}\right).
	\end{align}
\label{col1}}
\IEEEproof{
	Corollary \ref{col1} follows by letting $r_j\!=\!1, \forall j\!\in\! [J]$ and  $r^\prime_{j^\prime}\!=\!1, \forall j^\prime\!\in\! [J^\prime]$ in \eqref{Rk_us_multi}.}
\section{Beam Selection}
\label{sec3.3}
With the sum-rate expression \eqref{Rk_us_multi}, we are now in the position to present the beam selection scheme that aims to maximize the sum-rate. Specifically, for a given $\Gamma$, we select $\{\mathcal{G}_k\}$ to maximize \eqref{Rk_us_multi}, \ie
\begin{align}
\label{beamselection}
(\mathcal{G}_1,\cdots,\mathcal{G}_K)\! =\!\argmax_{\substack{\mathcal{G}_i \bigcap \mathcal{G}_j =\emptyset, \forall i,j \\\abs{\mathcal{G}_k}=\Gamma,\forall k}}\!\!\! R_{\text{sum}}(\mathcal{G}_1,\cdots,\mathcal{G}_K;1,2,\dots,K).
\end{align}
{\blue We show in Appendix \ref{appa0} that the beam selection problem in \eqref{beamselection} is NP-hard if $\Gamma \geq 3$, and is \emph{at least} as hard as the canonical maximal matching problem \cite{ComOptimization} if $\Gamma < 3$.
}
In both cases, the high complexity prohibits the evaluation of the sum-rate at all possible choices of $\{\mathcal{G}_k:k\in [K]\}$ in \eqref{Rk_us_multi}. To tackle this challenge, we propose two iterative algorithms for solving \eqref{beamselection} based on the forward stepwise (FS) searching and the Gibbs sampling. Furthermore, approximations for \eqref{Rk_us_multi} are introduced based on the structure of the sum-rate in the mediate and high SNR regimes for the proposed algorithms, while the simplification of the sum-rate expression in the low SNR regime is derived in Section \ref{lowsec}.
The approximations significantly reduce the computational complexity by eliminating the exponential integrals.
\subsection{FS-Based Beam Selection with a Simplified Sum-Rate Expression}\label{sec_FS}
The FS scheme searches $\{\mathcal{G}_k\}$ sequentially by activating one user per iteration {\blue in random order}. At iteration $k$, we choose
\begin{align}
\label{temp41}
\mathcal{G}_{k}=\argmax_{\substack{\abs{\mathcal{G}_k}=\Gamma\\\mathcal{G}_k\bigcap \mathcal{G}_i=\emptyset, i<k}}R_{\text{sum}}(\{\mathcal{G}_i:i <k\}, \mathcal{G}_{k};1,\dots,k).
\end{align}

As aforementioned, we seek for an approximation of \eqref{temp41} to reduce the searching complexity. 
First, by Jensen's inequality, $R_{\text{sum}}$ in \eqref{temp41} is upper bounded as
\begin{align}
\label{temp58}
R_{\text{sum}}&(\{\mathcal{G}_i:i <k\}, \mathcal{G}_{k};1,\dots,k)\leq \log_2(1+\E\left[\text{SINR}_k\right])+\sum_{i<k} \log_2(1+\E\left[\text{SINR}_i\right]).
\end{align}
Then, we approximate the r.h.s. of \eqref{temp58} as
\begin{align}
&\log_2(1+\E\left[\text{SINR}_k\right])+\sum_{i<k} \log_2(1+\E\left[\text{SINR}_i\right])\nonumber\\
\overset{(a)}\geq &\log_2 \left(1+\frac{\frac{\gamma}{\Upsilon\Gamma} \sum\limits_{l\in \mathcal{G}_k}\sigma_{k,l}^2}{1+\frac{\gamma}{\Upsilon\Gamma} \sum\limits_{l\in \Omega_{k}(k\!-\!1)\setminus\mathcal{G}_k}\sigma_{k,l}^2}\right)+\sum_{i <k} \log_2\left( 1+\frac{\frac{\gamma}{\Upsilon\Gamma}\sum\limits_{l\in \mathcal{G}_i} \sigma_{i,l}^2}{1+\frac{\gamma}{\Upsilon\Gamma}\left( \sum\limits_{l \in \Omega_i(k-1)\setminus\mathcal{G}_i}\sigma^2_{i,l}+\sum\limits_{l \in  \mathcal{S}_i\bigcap \mathcal{G}_k}\sigma^2_{i,l}\right) } \right)\nonumber\\
\overset{(b)}>& \log_2 \left(\frac{\sum\limits_{l\in \mathcal{G}_k}\sigma_{k,l}^2}{\frac{\Upsilon\Gamma}{\gamma}+ \sum\limits_{l\in \Omega_{k}(k\!-\!1)\setminus\mathcal{G}_k}\sigma_{k,l}^2}\right)+\sum_{i <k} \log_2\left( \frac{\sum\limits_{l\in \mathcal{G}_i} \sigma_{i,l}^2}{\frac{\Upsilon\Gamma}{\gamma}+ \sum\limits_{l \in \Omega_i(k-1)\setminus\mathcal{G}_i}\sigma^2_{i,l}+\sum\limits_{l \in  \mathcal{S}_i\bigcap \mathcal{G}_k}\sigma^2_{i,l} } \right)\nonumber\\
=&\log_2 \!\left( \sum_{l\in \mathcal{G}_k}\!\sigma_{k,l}^2\right)\!-\!\sum_{i <k}\log_2\!\left(\!\frac{\Upsilon\Gamma}{\gamma}\!+\! \sum\limits_{l \in \Omega_i(k-1)\setminus\mathcal{G}_i}\!\!\sigma^2_{i,l}\!+\!\sum\limits_{l \in  \mathcal{S}_i\bigcap \mathcal{G}_k}\!\!\sigma^2_{i,l}\right)+\psi_k
\nonumber\\
\overset{(c)}\geq&\log_2 \left( \sum_{l\in \mathcal{G}_k}\sigma_{k,l}^2\right)-\log_2\left(\frac{\Upsilon\Gamma}{\gamma}+\sum_{i <k}\left(  \sum\limits_{l \in \Omega_i(k-1)\setminus\mathcal{G}_i}\sigma^2_{i,l}+\sum_{l \in  \mathcal{S}_i\bigcap \mathcal{G}_k}\sigma^2_{i,l}\right) \right)+\psi_k\nonumber\\
\triangleq& R_{\text{sum}}^{\text{approx}}\left(\{\mathcal{G}_i:i <k\}, \mathcal{G}_{k};1,\dots,k\right ),
\label{approsum-rate}
\end{align}
where $\Omega_{i}(t)$ is the active beam set of user $i$ at iteration $t$; {\blue $(a)$ is from Mullen's inequality \cite{MullenInequlity}}; $(b)$ is from $\log_2(1+x)>\log_2(x)$ for $x>0$; $(c)$ is from the convexity of $-\log_2(c+x)$ for some positive constant $c$; and $\psi_k\triangleq-\log_2\left( \frac{\Upsilon\Gamma}{\gamma}+ \sum_{l\in \Omega_{k}(k\!-\!1)\setminus\mathcal{G}_k}\sigma_{k,l}^2\right)$ $+\sum_{i <k}\log_2\left(\sum_{l\in \mathcal{G}_i} \sigma_{i,l}^2\right)$ represents the term independent of the choice of $\mathcal{G}_k$.

As discussed later in this subsection, the FS scheme aims to simultaneously increase the user signal power and decrease the interference. This implies that the inequalities in  \eqref{temp58} and \eqref{approsum-rate} are tight when the system does not work in the noise-dominated regime. Therefore, by substituting  \eqref{approsum-rate} into \eqref{temp58}, in the mediate to high SNR regime we obtain
\begin{align}
\label{temp59}
R_{\text{sum}}&(\{\mathcal{G}_i:i <k\}, \mathcal{G}_{k};1,\dots,k) \approx R_{\text{sum}}^{\text{approx}}\left(\{\mathcal{G}_i:i <k\}, \mathcal{G}_{k};1,\dots,k\right ).
\end{align}
Verifications of \eqref{temp59} can be found in Section \ref{sec_validation}.
With \eqref{temp59}, the optimization problem \eqref{temp41} can be simplified as
\begin{align}
\mathcal{G}_{k}&=\argmax_{\substack{\abs{\mathcal{G}_k}=\Gamma\\\mathcal{G}_k\bigcap \mathcal{G}_i=\emptyset, i<k}} R_{\text{sum}}^{\text{approx}}\left(\{\mathcal{G}_i:i <k\}, \mathcal{G}_{k};1,\dots,k\right ) \nonumber
\end{align}
\begin{align}
&=\argmax_{\substack{\abs{\mathcal{G}_k}=\Gamma\\\mathcal{G}_k\bigcap \mathcal{G}_i=\emptyset, i<k}}\log_2 \left( \sum_{l\in \mathcal{G}_k}\sigma_{k,l}^2\right)-\log_2\left(\frac{\Upsilon\Gamma}{\gamma}+\sum_{i <k}\left(  \sum\limits_{l \in \Omega_i(k-1)\setminus\mathcal{G}_i}\sigma^2_{i,l}+\sum_{l \in  \mathcal{S}_i\bigcap \mathcal{G}_k}\sigma^2_{i,l}\right) \right)\nonumber\\
&=\argmax_{\substack{\abs{\mathcal{G}_k}=\Gamma\\\mathcal{G}_k\bigcap \mathcal{G}_i=\emptyset, i<k}}\frac{\frac{\gamma}{\Upsilon\Gamma}\sum_{l\in \mathcal{G}_k}\sigma_{k,l}^2}{1+\frac{\gamma}{\Upsilon\Gamma}\left(\sum_{i <k} \sum\limits_{l \in \Omega_i(k-1)\setminus\mathcal{G}_i}\sigma^2_{i,l}+\sum_{i <k}\sum_{l \in  \mathcal{S}_i\bigcap \mathcal{G}_k}\sigma^2_{i,l}\right) }\label{temp42}.
\end{align}
%
Intuitively, the numerator in \eqref{temp42} represents the signal power of the newly activated user $k$, contributing to the achievable rate gain at iteration $k$. Meanwhile, the denominator in \eqref{temp42} represents the noise and interference power at iteration $k$. Therefore, optimizing over \eqref{temp42} matches our common sense: It maximizes the signal power of the new user (and hence maximizes its achievable rate), while keeping the overall interference at a low level.
\subsection{Gibbs-Sampling-Based Beam Selection}\label{sec_gibbs}
The performance of the FS-based beam selection is compromised by severe interference when $K$ is large. To see this, recall that \eqref{temp42} only involves the interference for users $1,2,\dots,k$ at iteration $k$. The currently inactive users are not considered, since the FS scheme activates users sequentially. This greedy approach generally suffers from a certain performance loss. To reduce this loss, we employ the idea of Gibbs sampling \cite{PowerControl_Qian} to select beams, which avoids the high computational complexity of the combinatorial search.

Specifically, Gibbs sampling sequentially updates $\mathcal{G}_k$ at step $i \in [I]$ according to the probability distribution
\begin{align}
\label{temp53}
\Lambda_i(\mathcal{G}_k|\mathcal{G}_{-k})=\frac{\exp\left(-\beta_i^{-1}/J(\mathcal{G}_k;\mathcal{G}_{-k}) \right) }{\sum_{\substack{\abs{\tilde{\mathcal{G}}_k}=\Gamma\\\tilde{\mathcal{G}}_k\subset [L]\setminus\mathcal{G}_{-k}}}\exp\left(-\beta_i^{-1}/J(\tilde{\mathcal{G}}_k;\mathcal{G}_{-k}) \right)},
\end{align}
where $\mathcal{G}_{-k}=\bigcup_{i \neq k} \mathcal{G}_i$; $\beta_i>0$ denotes the ``temperature” parameter; and $J(\cdot)$ is the objective to be maximized (\ie the sum-rate). Similarly to \eqref{temp42}, we simplify the sum-rate expression by using the same approximation in \eqref{temp59}, yielding
\begin{align}
J(\mathcal{G}_k;\mathcal{G}_{-k})=\frac{\sum_{l\in \mathcal{G}_k}\sigma_{k,l}^2}{{\Upsilon\Gamma}/{\gamma}+\sum_{i \neq k} \sum_{l \in \Omega_i\setminus\mathcal{G}_i}\sigma^2_{i,l}+\sum_{i \neq k}\sum_{l \in  \mathcal{S}_i\bigcap \mathcal{G}_k}\sigma^2_{i,l}}.
\end{align}
To accelerate the convergence of the proposed algorithm, several modifications are introduced as follows.
\begin{itemize}
	\item Since the value of $J(\cdot)$ may vary significantly from one iteration to another, we normalize $J(\cdot)$ before computing \eqref{temp53} to stabilize the update, \ie we compute
\begin{align}
\tilde{J}(\mathcal{G}_k;\mathcal{G}_{-k})=J(\mathcal{G}_k;\mathcal{G}_{-k})/\max\limits_{\substack{\abs{\tilde{\mathcal{G}}_k}=\Gamma\\\tilde{\mathcal{G}}_k\subset [L]\setminus\mathcal{G}_{-k}}}J(\tilde{\mathcal{G}}_k;\mathcal{G}_{-k}),
\end{align}
and replace ${J}(\cdot)$ with $\tilde{J}(\cdot)$ in \eqref{temp53}.
\item We employ a slowly decreasing ``cooling schedule” \cite{Annealing} for $\beta_i$: $\beta_{i\!+\!1}=\rho\beta_i$ for $0<\rho<1$. In other words, we encourage the sampler to explore the searching space at the beginning period and force convergence as the iteration proceeds. The exploration-exploitation trade-off is controlled by the cooling rate $\rho$.
\item It is possible that the sampling space is multimodal and has a low transition probability at the connections between the modes. As a consequence, the sampler is easy to get stuck in a stationary point with a small $\beta_i$. We overcome this problem by recording the best sample during iterations, \ie keep tracking the best result in the exploration stage. Furthermore, to encourage the sampler to explore various modes along multiple paths, the Gibbs sampler is invoked three times independently and the best result is picked as the final choice.
\end{itemize}
\section{Sum-Rate Characterization}\label{sec_asymoptotic}
In Section \ref{sec_sumrate}, a sum-rate expression for the proposed massive MIMO system is presented. In this section, we carry out quantitative studies on the sum-rate performance of the proposed algorithms. The first two ingredients in our analysis are the asymptotic results in the low and high SNR regimes.
\subsection{Low-SNR Analysis}
\label{lowsec}
\begin{proposition}
	When $\gamma \to 0$, the sum-rate  converges to  
	\begin{align}
	\label{lowsnr}
	\lim_{\gamma\to 0}\frac{R_{\text{sum}}}{\gamma}=	\frac{\sum_{k=1}^K\sum_{l\in \mathcal{G}_k}\sigma^2_{k,l}}{\Upsilon\Gamma \ln 2}.
	\end{align}
	\label{pro2}
\end{proposition}
\IEEEproof{See Appendix \ref{appb}.}

In the low-SNR regime, the sum-rate linearly scales with $\gamma$ due to the domination of the noise. According to Proposition \ref{pro2}, the users can select the beams $\{\mathcal{G}_k\}$ aggressively and independently to maximize their own received signal power, which leads to a simple beam selection principle in the low SNR regime.

\subsection{High-SNR Analysis}
\label{highsec}
\begin{proposition}
	Under the same assumption as in Corollary \ref{col1}, when $\{\mathcal{G}_k\}$ is given and $\gamma$ is large, the sum-rate is given by
	\begin{align}
	\label{highsnr}
R_{\text{sum}}\approx C_1 \log_2(\gamma)+C_2,
	\end{align}
	\label{pro3}
	where $C_1$ and $C_2$ are constants dependent of $\{\mathcal{G}_k: k \in [K]\}$.
\end{proposition}
\IEEEproof{See Appendix \ref{appc}.}

 Although Proposition \ref{pro3} only applies to the case that $\{\sigma_{k,l}, l \in \Omega_k\}$ are distinct, it reveals the general relationship between the sum-rate and the transmission power. Note that the linear scaling of $R_{\text{sum}}$ w.r.t. $\log_2(\gamma)$ at the high-SNR regime coincides with the definitions in \eqref{temp64} and \eqref{temp57}.
\subsection{Effect of $\Gamma$}\label{sec_Gamma}
We now turn to the study on the parameter $\Gamma$. On one hand, even though $\Gamma$ can be any positive integer, we usually implement a small $\Gamma$ for the following reasons. First, since $\abs{\mathcal{S}_k}\ll L$, the number of effective paths is very limited and a large $\Gamma$ is wasteful. Second, $\Gamma$ is constrained by $\Gamma < L/K$  to satisfy $\mathcal{G}_i \bigcap \mathcal{G}_j =\emptyset$. Third, for fixed transmission power $P$, a factor $1/\Gamma$ exists in computing the SINR in \eqref{SINR_Gamma2}. This has a negative effect on the sum-rate \eqref{Rk_us_multi}, since the function $\exp(\cdot)E_n(\cdot)$ is decreasing w.r.t. its argument. Thus, a larger $\Gamma$ does not necessarily imply better performance. Four, the rate of the BTBC is usually less than $1$ when $\Gamma >2$,  causing a significant sum-rate loss. {\blue Finally, for a large $\Gamma$, since multiple beams are utilized for exploiting the diversity gain, fewer degrees of freedom are available in beam selection for eliminating interference.} 

On the other hand, the system may experience a temporary link failure especially when $\Gamma=1$, because of the strong destruction of $s_{k,l}$, a.k.a. a deep fade. Choosing a relatively larger $\Gamma$, if allowed, significantly reduces the probability that a deep fade occurs, as it provides a higher diversity gain with the proposed BTBC.

We emphasize that the choice of $\Gamma$ is closely related to the system size. When $N/K$ is large, the overlaps of the available beams between users are generally neglectable. A relatively larger $\Gamma$ is preferable for achieving a higher diversity gain without causing severe interference. On the contrary, we need a conservative choice of $\Gamma$ to suppress the IUI when $N/K$ is small.

To sum up, we generally prefer a small number of beams to be selected in the proposed algorithm. We recommend  $\Gamma=1$ when $N/K$ is small due to the interference mitigation concerns. Otherwise, we recommend  $\Gamma=2$ when $N/K$ is large.
\subsection{Effect of $K$}\label{sec_K}
Similarly to the analysis in the previous subsection, the relationship between $R_{\text{sum}}$ and $K$ is not monotonic. Starting from a small group of activated users, adding a new user (\ie enlarging $K$) generates more summation terms on the r.h.s. of \eqref{temp57}, thereby increasing the sum-rate performance of the system. However, increasing $K$ yields a larger $\Omega_{i}$ for the previously activated user $i$. This may lead to a decrease in $R_i$ since more interference is introduced, especially in the high SNR regime. As a result, the sum-rate first increases and then may decrease as $K$ becomes large. {\blue Moreover, we note that beams should be sufficiently utilized to provide multiuser access rather than exploiting the multiple path diversity when K is large, leaving less space for beam selection in our design. }
\section{Numerical Results}\label{sec_simulation}
\subsection{Simulation Setups}
In this section, we present the numerical results for the proposed beamforming algorithms, with the FS-based beam selection and the Gibbs-sampling-based beam selection. Unless otherwise specified, the simulation setting is given as follows. We assume that the BS is equipped with a half-wavelength ULA, \ie $fd/c_0=0.5$ in \eqref{ULA}. We further assume that the channel is invariant with the coherence duration $T\!=\!100$. We set $N\!=\!L\!=\!64$, $P(k)=5$ for $\forall k$. {\blue The AoDs of each user are randomly selected from the length-$L$ set $\thetav$. Denote the $l$-th element of  $\thetav$ by  $\theta_l$. We generate $\theta_l$ such that  $\sin\theta_l=\frac{2l-1-L}{L}+\kappa_l$ with $\kappa_l$ uniformly drawn from $[-\frac{1}{2L},\frac{1}{2L}]$. With the choice of $\thetav$, we ensure that $\{\sin\theta_l\}$ covers the whole range of $[-1,1]$, and we control the randomness in the angles by $\kappa_l$. Moreover, we make sure that the angular separation between any pair of nearby angles satisfies $\frac{1}{L}\leq \sin\theta_{l+1}-\sin\theta_{l}\leq \frac{3}{2L}$. }
$\{s_{k,p}\}$ are i.i.d. and are drawn from $\CN(0,1)$. $\{\varsigma^2_{k,p}\}$ are i.i.d. and are uniformly drawn from $[0.1,1]$ and normalized to have a unit-sum for $\forall k$. For the Gibbs-sampling-based beam selection, we set $I=5K$, $\beta_1=0.1$, and $\rho=0.95$. All simulation results are obtained by averaging over $300$ Monte Carlo trials, unless otherwise specified. 

The following state-of-the-art SBF algorithms are taken as baselines. Both of the baseline algorithms design the beamformer $\Vv$ without utilizing instantaneous CSIT. 
\begin{itemize}
\item Baseline 1 \cite{SBF1}: The beamforming vector $\vv_k$ is designed to maximize a lower bound of the signal-to-leakage-noise ratio (SLNR). Specifically,
\begin{align}
\vv_k=\uv_{\max},
\end{align}
where $\uv_{\max}$ is the normalized principle eigenvector of $\left\{\left( \frac{1}{\gamma}\Iv_{N}+\sum_{j\neq k} \Rv_j(f)\right)^{-1} \Rv_k(f) \right\}$. To be consistent with the proposed framework, we compute $ \Rv_k(f)= \Av(\varthetav,f) \Sigmav_k\Av(\varthetav,f) ^H$.
\item Baseline 2 \cite{SBF3_2}: The DFT basis $\Fv=[\fv_1,\dots,\fv_N]$ is used to compute $\Gv_k=\Fv^H \Rv_k(f)\Fv$, where $[\Fv]_{nm}=e^{-j\pi (n-1)(m-1)/N}/\sqrt{N}$. Then, a similar SLNR maximization criteria yields the beamforming vector as $\vv_k=\fv_{n(k)}$, where
 \begin{align}
 n(k)=\argmax_{n \in [N]} \frac{g_{k,n}}{1/\gamma+\sum_{j\neq k}g_{j,n}},
 \end{align}
 where $g_{k,n}$ is the $n$-th diagonal entry of $\Gv_k$.
\end{itemize}
Besides, we also include the following training-based beamforming algorithms for comparisons, where the closed-loop downlink channel training is used for CSIT acquisition.
{\blue
\begin{itemize}
	\item Zero-forcing beamforming (ZFBF): Orthogonal training sequences with length $N$ are broadcast to the users. For user $k$, the collection of the received signals, after projecting over the orthogonal training sequence, is given by $\hat \hv_k=\hv_k+\nv_k^\prime$ with $\nv_k^\prime \sim \CN({\bf 0},1/\gamma \Iv_N )$. All the users feed back the information on $\{\hat \hv_k\}$ to the BS\footnote{We neglect the overhead and the additional errors in the feedback process for simplicity.}. Based on that, the BS computes the beamforming vectors as 
	\begin{align}
		\vv_k =\frac{ \Pv_k\hat\hv_k}{\norm{\Pv_k\hat \hv_k}_2}, \forall k,
	\end{align}
	where $\Pv_k=\Iv_N-\hat \Hv_{-k}(\hat \Hv_{-k}^H\hat \Hv_{-k})^{-1}\hat \Hv_{-k}^H$ with $\hat \Hv_{-k}\triangleq[\hat \hv_1,\dots,\hat \hv_{k-1},\hat \hv_{k+1},\dots,\hat \hv_{K}]$. Taking the training overhead into account, the sum-rate is given by 
	\begin{align}
	R_{\text{sum}}=\left(1-\frac{N}{T}\right)\sum_{k=1}^K \E\left[\log_2 \left( 1+\text{SINR}_k\right) \right].
	\end{align}
	\item Two-stage beamforming (TSBF) \cite{MIMO_GCaire}: The uplink channel sample covariance matrices are computed each with $1000$ observations. The uplink CCMs are obtained by projecting the sample covariance matrices to the Toeplitz, positive semidefinate cone. Then, the downlink CCMs are determined by an extrapolation scheme, where we set $f^{\text{ul}}/f=\frac{1950}{2140}$ \cite{MIMO_GCaire}.  Finally, the outer precode is designed by solving a mixed integer linear program and the inner precoder is computed as ZFBF discussed above.
\end{itemize}
}

\subsection{Validation of Sum-Rate Expressions}\label{sec_validation}
\begin{figure}[!t]
	\centering
	\includegraphics[width=3.4 in]{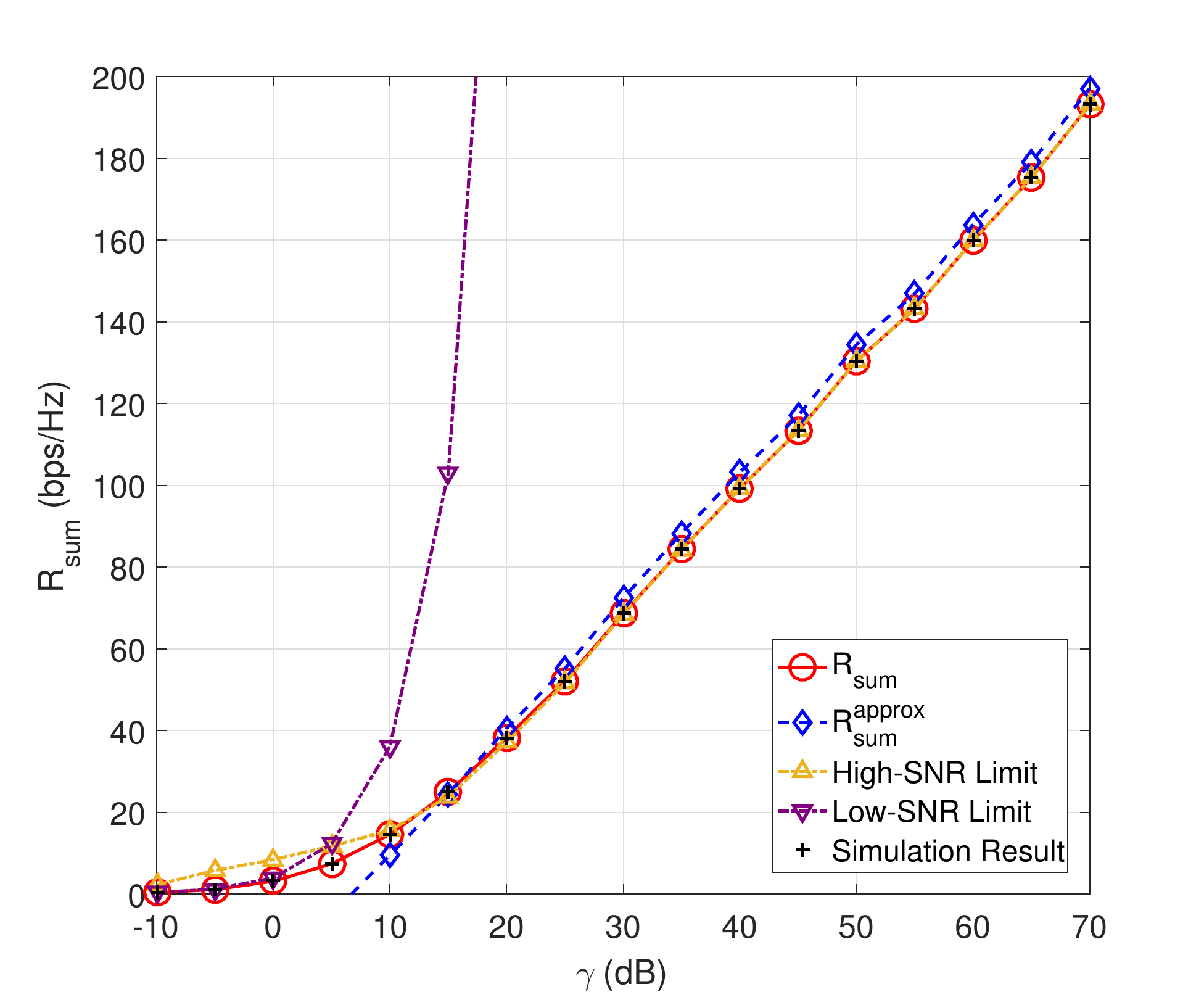}
	\caption{Sum-rate performance evaluated by different expressions at $K=10$, where the FS-based beam selection with $\Gamma=2$ is adopted.}
	\label{bound}
\end{figure}
In this subsection, we verify the sum-rate expression \eqref{Rk_us_multi}, its asymptotic limits \eqref{lowsnr}--\eqref{highsnr}, and its approximation \eqref{temp59}. In Fig. \ref{bound}, the results obtained by the FS-based beam selection with $\Gamma=2$ are evaluated with these metrics. It can be seen that the sum-rate expression \eqref{Rk_us_multi} exactly matches the simulation result in the entire SNR regime. Meanwhile, the low- and high-SNR limits are identical to $R_{\text{sum}}$ in the low and high SNR regimes, respectively. As an approximation,  $R_{\text{sum}}^{\text{approx}}$ fits the true curve in the mediate and high SNR regimes, but has a poor behavior when $\gamma<10\text{ dB}$. This has been previously discussed in Section \ref{sec_FS}.
Fortunately, the low-SNR limit \eqref{lowsnr} can be used as a good approximation of \eqref{beamselection} when $\gamma<10\text{ dB}$.

Besides, the simulation reveals the relationship between $R_{\text{sum}}$ and $\gamma$: $R_{\text{sum}}$ grows linearly as $\gamma$ increases in the low SNR regime, and scales up linearly w.r.t. the logarithm of $\gamma$ in the high SNR regime. This observation coincides with the analysis in Section \ref{sec_asymoptotic}.

\subsection{Comparisons With Exhaustive-Search-Based Beam Selection}
	\begin{figure}[!t]
	\begin{minipage}[t]{0.5\linewidth}
		\centering
		\includegraphics[width=3.5 in]{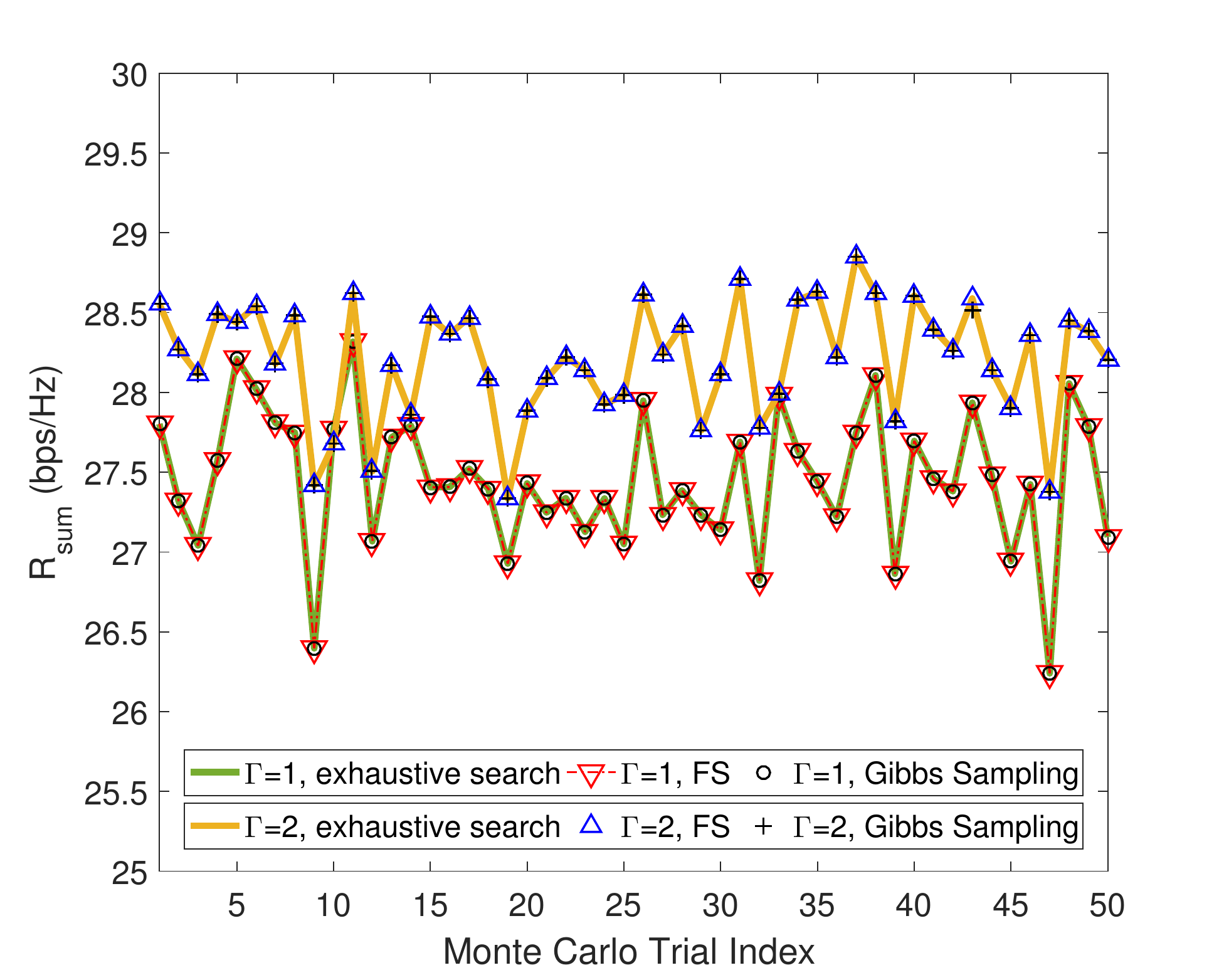}
		\caption{Sum-rate performance for various schemes over $50$\protect\\ Monte Carlo Trials at $P=40$ dB and $K=3$.}
		\label{opt_sum}
	\end{minipage}%
	\begin{minipage}[t]{0.5\linewidth}
		\centering
		\includegraphics[width=3.5in]{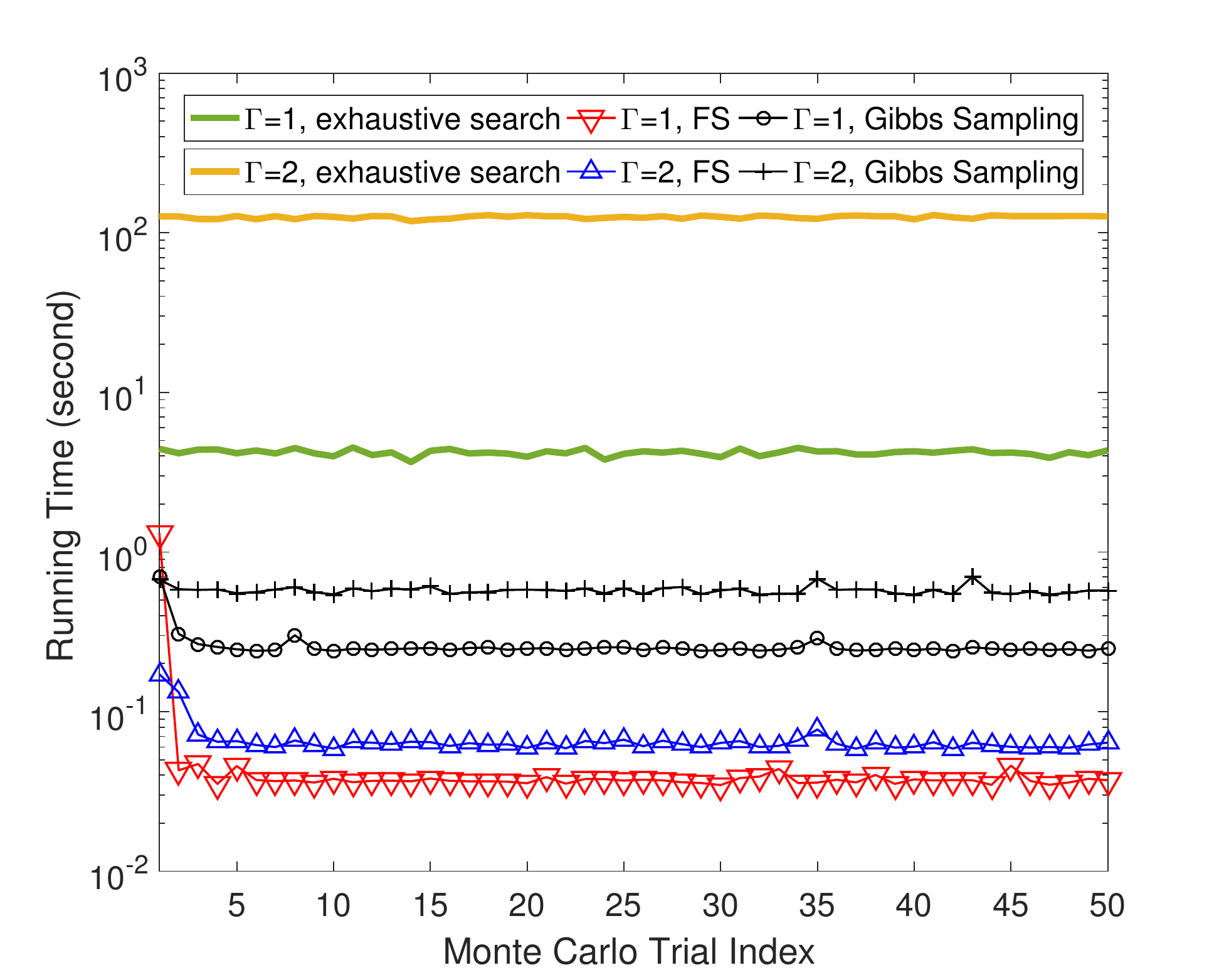}
		\caption{Running time over $50$ Monte Carlo Trials at $P=40$ dB and $K=3$.}
		\label{opt_time}
	\end{minipage}
\end{figure}
We investigate the optimality of the two proposed beam selection schemes. We conduct a simulation that compares the solutions from the proposed schemes and the optimal $\{\mathcal{G}_k\}$ obtained from an exhaustive search method over \eqref{beamselection}. The simulation platform is a Dell Optiplex $9010$ Desktop with an $i7\!-\!3770/3.40$GHz Quad-Core CPU and $16$GB RAM. As discussed in Section \ref{sec3.3}, the complexity of the exhaustive search grows exponentially with the system parameters such as $K$ and $\Gamma$. For example, it takes more than an hour to finish one exhaustive search trial for $K=4$ and $\Gamma=2$. Thus, we restrict our simulation to the case of $K=3$. Figs. \ref{opt_sum} and \ref{opt_time} plot the sum-rate performance and the running time for different selection schemes. Under this setting, both of the two proposed schemes achieve the optimal performance for all $50$ trials. Moreover, the running time of the proposed schemes is at least one order of magnitude lower than that of the exhaustive search, verifying the effectiveness of the two schemes. We note that the proposed beam selection schemes do not necessarily guarantee optimality in general, especially when $K$ is large. However, it is difficult to provide verifications for a large $K$ due to the overwhelming complexity of the exhaustive search.
	
\subsection{Simulations Under Perfect Spatial Information}
	\begin{figure}[!t]
	\begin{minipage}[t]{0.5\linewidth}
	\centering
	\includegraphics[width=3.5 in]{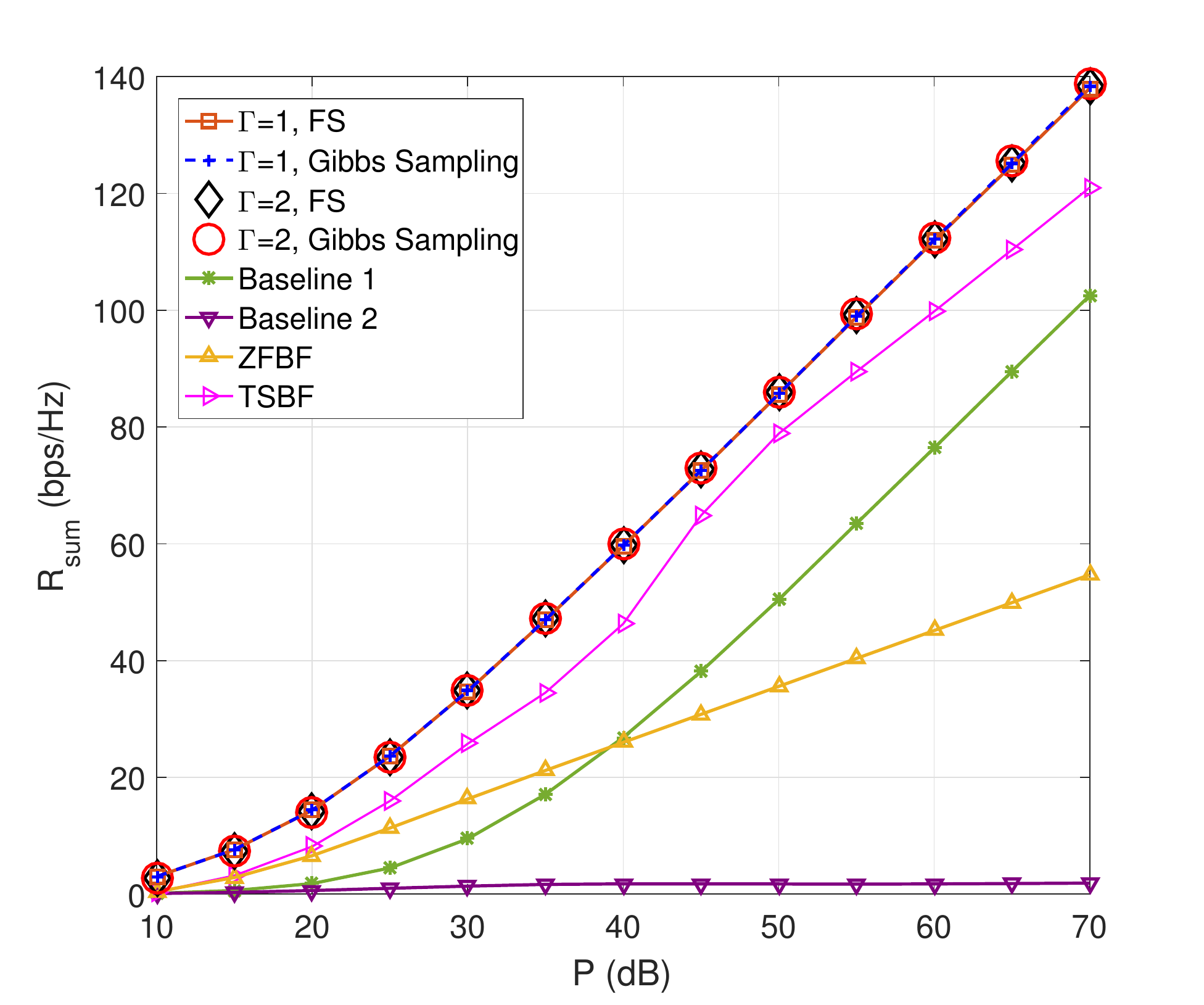}
	\caption{Sum-rate versus total transmission power for $K=8$\protect\\ and $N=64$.}
	\label{P_R_sum}
	\end{minipage}%
	\begin{minipage}[t]{0.5\linewidth}
	\centering
	\includegraphics[width=3.5 in]{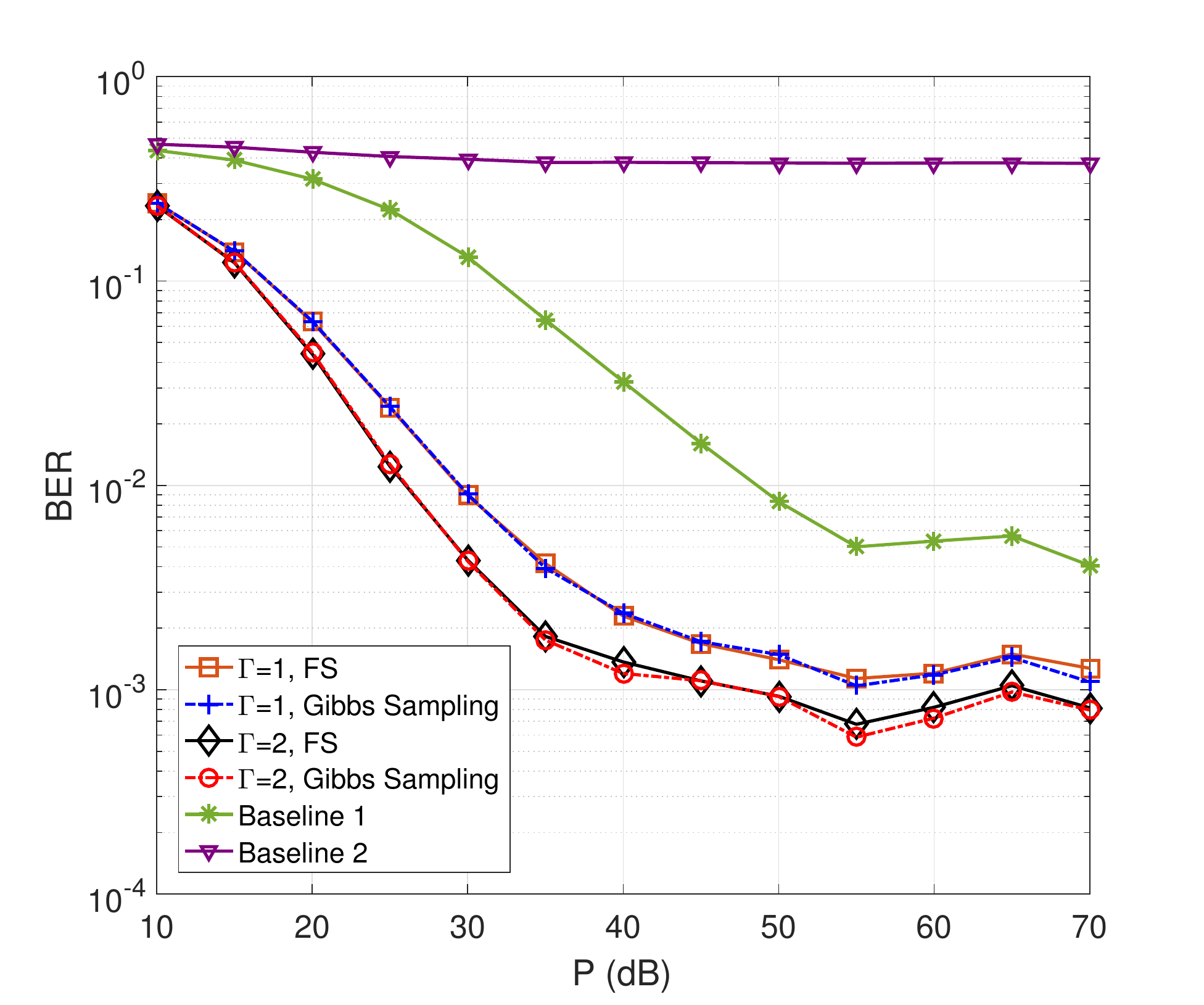}
	\caption{BPSK BER versus $P$ for $K=8$ and $N=64$.}
	\label{P_BER}
	\end{minipage}
	\end{figure}
In this section, the performance of the proposed beamforming algorithm is evaluated by comparing with the baselines.  Figs. \ref{P_R_sum} and \ref{P_BER} show the sum-rates and the bit error rates (BERs) with binary phase-shift keying (BPSK) at different transmission power levels. It can be seen that 1) the sum-rate results of all algorithms increase as $P$ increases and the proposed methods outperform the baselines at all energy levels; 2) as discussed in Section \ref{sec_Gamma}, the sum-rates for the proposed algorithms with $\Gamma=1$ and $\Gamma=2$ are almost the same;
3) although a larger $\Gamma$ does not necessarily imply a larger sum-rate, it improves the BER performance with a higher diversity gain; see the discussion in Section \ref{sec_Gamma}.
	\begin{figure}[!t]
	\begin{minipage}[t]{0.5\linewidth}
		\centering
		\includegraphics[width=3.5 in]{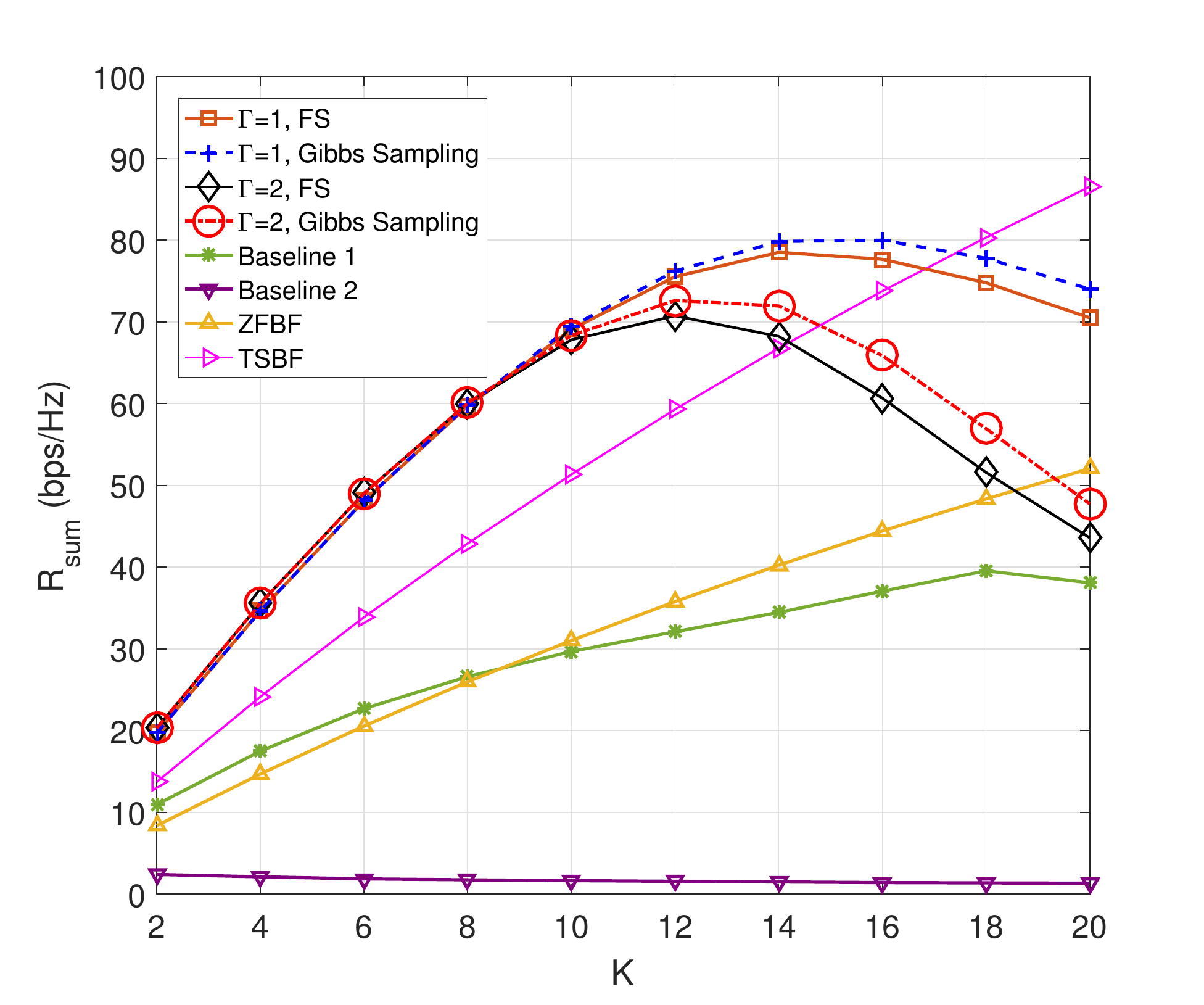}
		\caption{Sum-rate versus $K$ for $P=40$ dB and $N=64$.}
		\label{K_R_sum}
	\end{minipage}%
	\begin{minipage}[t]{0.5\linewidth}
		\centering
		\includegraphics[width=3.5 in]{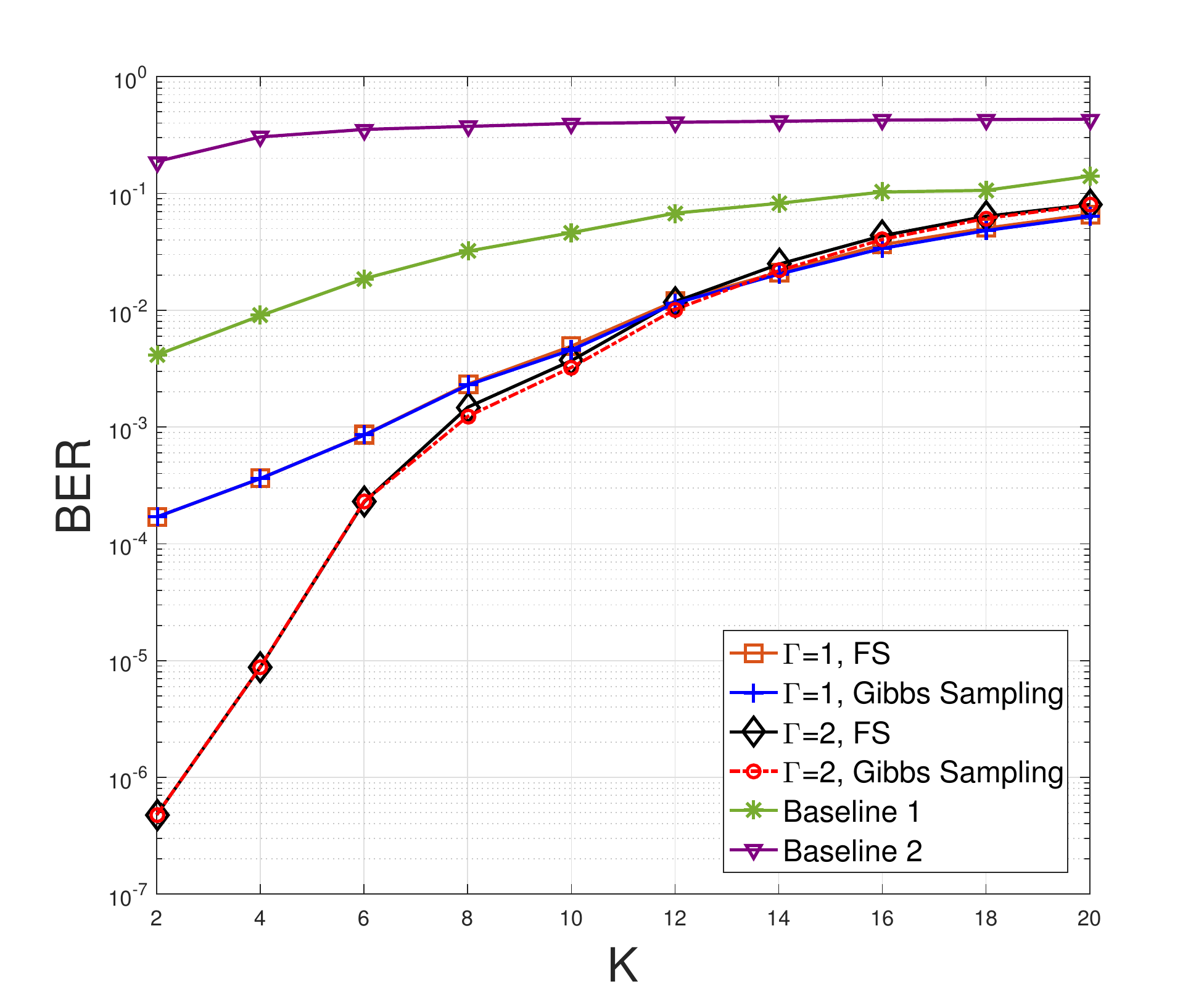}
		\caption{BER versus $K$ for $P=40$ dB and $N=64$.}
		\label{K_BER}
	\end{minipage}
\end{figure}
	\begin{figure}[!t]
	\begin{minipage}[t]{0.5\linewidth}
		\centering
		\includegraphics[width=3.5 in]{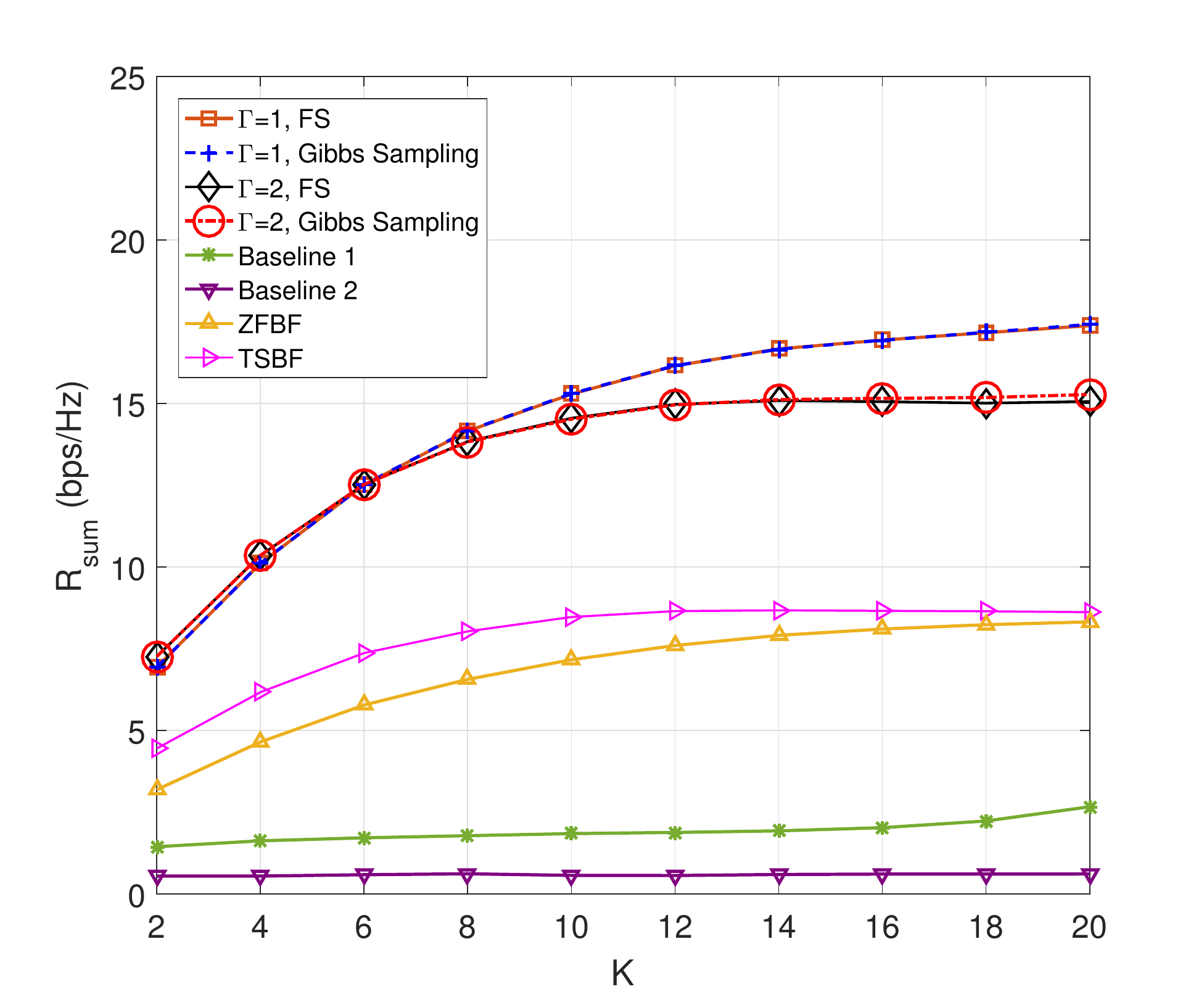}
		\caption{Sum-rate versus $K$ for $P=20$ dB and $N=64$.}
		\label{K_R_sum2}
	\end{minipage}%
	\begin{minipage}[t]{0.5\linewidth}
		\centering
		\includegraphics[width=3.5 in]{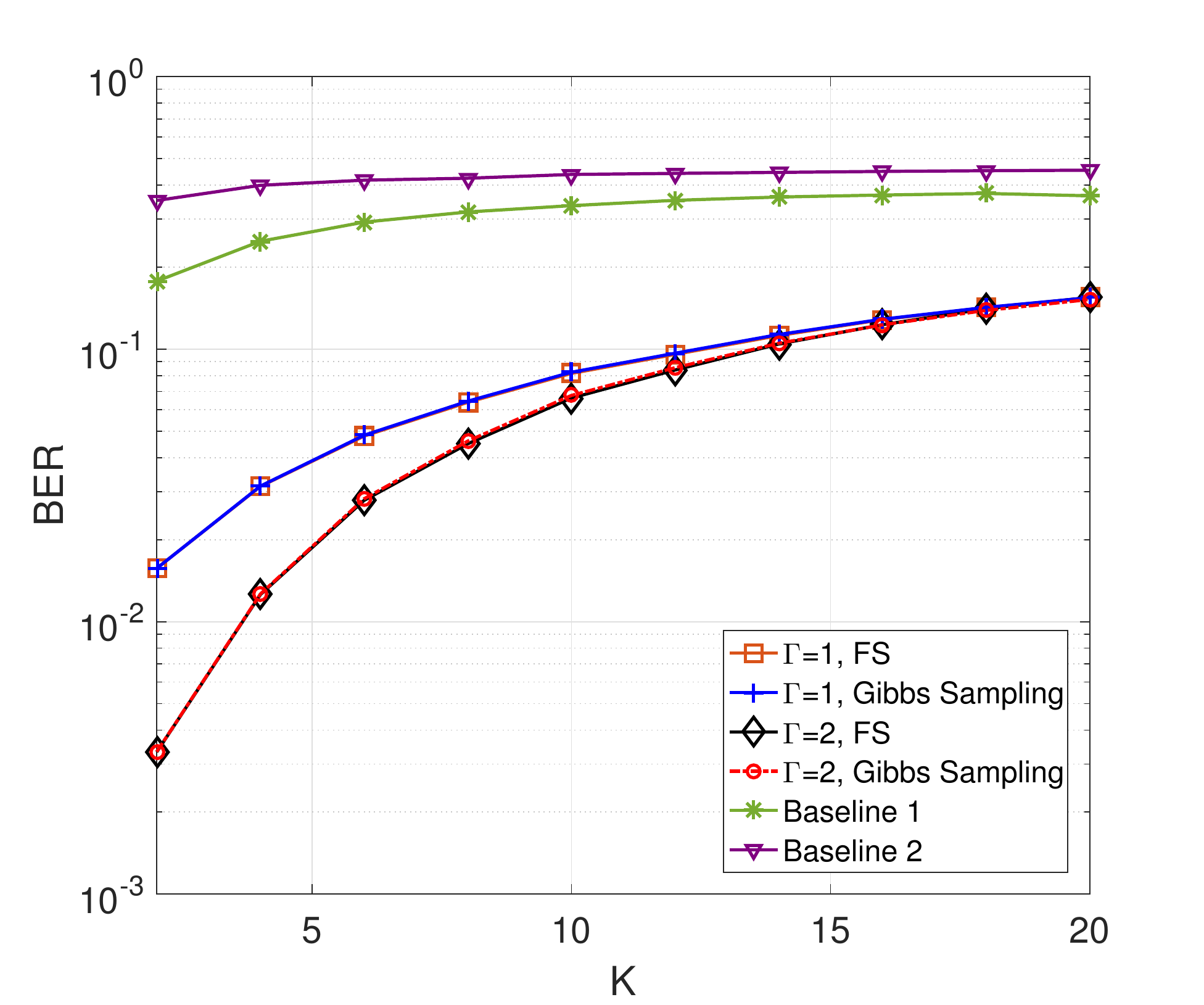}
		\caption{BER versus $K$ for $P=20$ dB and $N=64$.}
		\label{K_BER2}
	\end{minipage}
\end{figure}

We next investigate the effect of the number of users on the proposed algorithms. {\blue We plot the sum-rate and the BER performance versus $K$ with $P= 40$ dB in Figs. \ref{K_R_sum} and \ref{K_BER} and with  $P= 20$ dB in Figs. \ref{K_R_sum2} and \ref{K_BER2}.} We have the following observations. 1) As discussed in Section \ref{sec_K}, increasing $K$ results in a peak of $R_{\text{sum}}$ at a certain value of $K$ in the high SNR regime ($P=40$ dB). For  $P=20$ dB, since the noise dominates the IUI, we do not observe the decrease of the sum-rate with a large $K$. 2) For $P=20$ dB, our algorithms outperform the baselines in terms of both the sum-rate and the BER.  For $P=40$ dB, the proposed algorithms achieve better performance compared with  TSBF when $K<14$. For $K>14$, TSBF achieves a larger sum-rate as the proposed methods suffer from severe IUI; see the discussion in Section \ref{sec_K}.  3) In terms of the sum-rate, $\Gamma=1$ is preferable when $K>8$, as it alleviates severe interference. On the contrary, $\Gamma=2$ provides much higher diversity gain and reduces the error rate in a relatively small system with $K<8$; see the discussion in Section \ref{sec_Gamma}. 4) As mentioned in Section \ref{sec_gibbs}, Gibbs sampling improves the beam selection performance by alleviating IUI, especially when $K$ is relatively large.

\subsection{Simulations Under Imperfect Spatial Information}\label{sec_simulation2}
\begin{figure}[!t]
	\begin{minipage}[t]{0.5\linewidth}
		\centering
		\includegraphics[width=3.5 in]{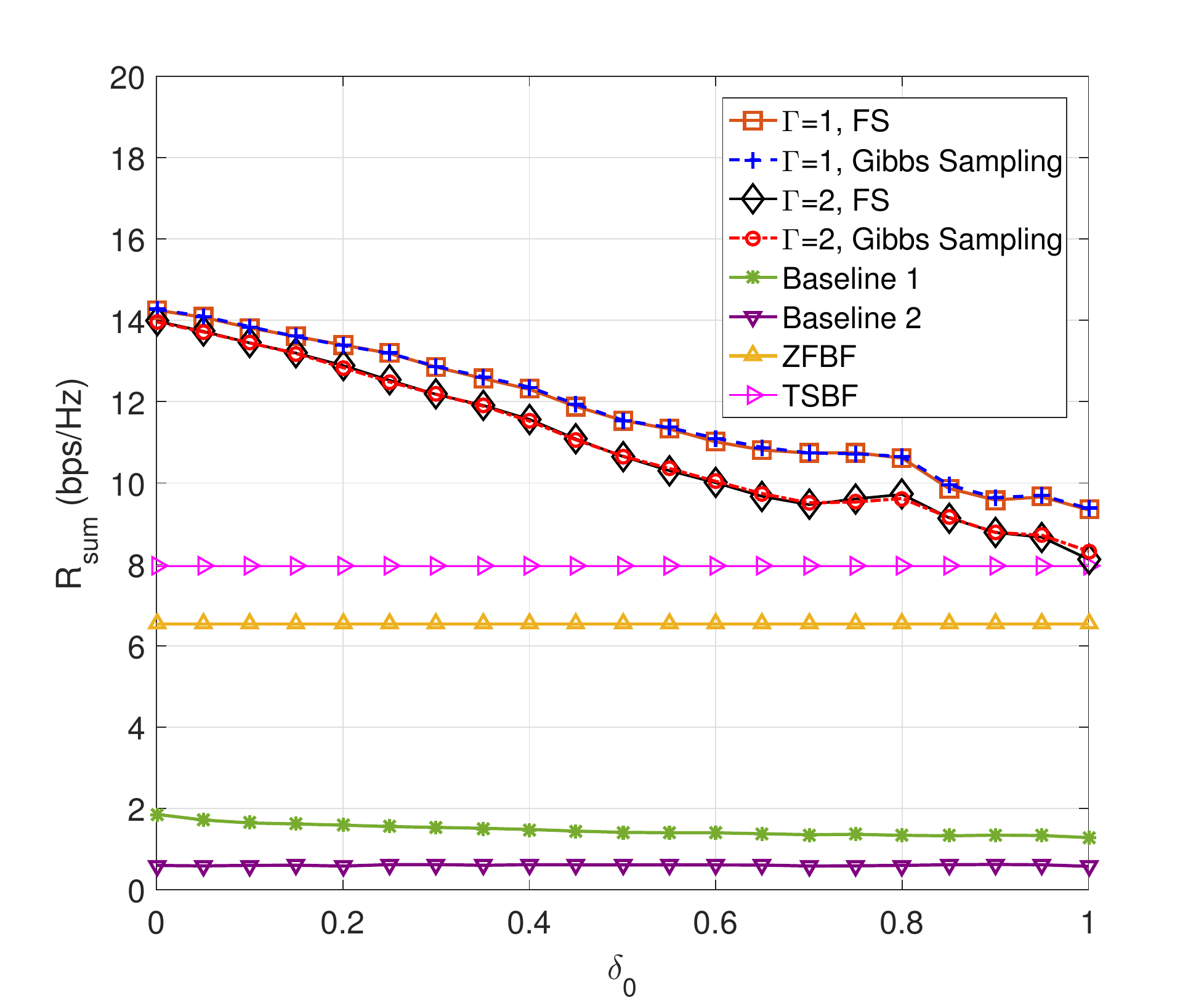}
		\caption{Performance for different angle mismatch levels\protect\\ at $P=20$ dB and $K=8$.}
		\label{fig_delta}
	\end{minipage}%
	\begin{minipage}[t]{0.5\linewidth}
		\centering
		\includegraphics[width=3.5 in]{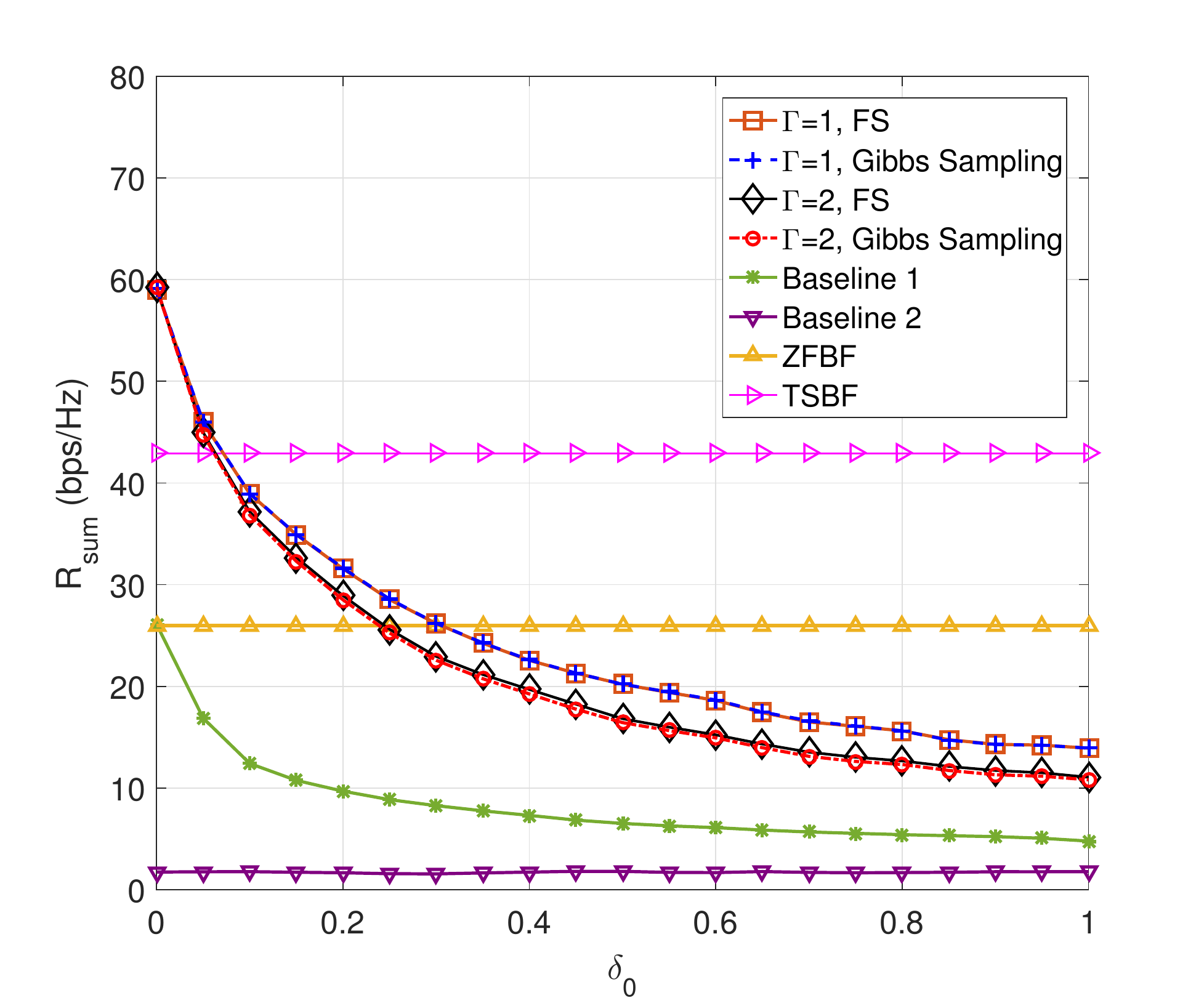}
		\caption{Performance for different angle mismatch levels at $P=40$ dB and $K=8$.}
		\label{fig_delta2}
	\end{minipage}
\end{figure}
We study the effect of the imperfect spatial information on the proposed algorithm. We start with the case of $\deltav\neq \bf{0}$ in \eqref{offgridmismatch}. {\blue Specifically, $\deltav$ is modelled as a Gaussian random vector following the distribution $\Norm({\bf 0},\delta_0 \Iv_{L})$ with a varying $\delta_0$ to control the mismatch level.  Figs. \ref{fig_delta} and  \ref{fig_delta2} plot the sum-rates of the proposed algorithms under different angle mismatch levels with $P=40$ dB and $P=20$ dB, respectively. } On one hand, ZFBF and TSBF are designed without utilizing the angle information. They are not affected by the angle mismatch. On the other hand, the performance decreases as $\delta_0$ increases for all schemes that depend on the angle information, including the proposed methods. {\blue We also see that the performance degradation becomes severer when the total transmission power is higher. This is because the angle mismatch problem introduces additional IUI, which has a negative effect on the achievable rate \eqref{temp64}, especially in the high SNR regime.}

{\blue To alleviate severe angle mismatch, we have to decrease the number of users $K$, so as to enlarge the angular separation between users. Fig. \ref{fig_rev} plots the performance of the proposed method with the FS-based beam selection scheme under different angle mismatch levels, where $\Gamma=1$ and $K=2, 4, 8$. We define the evaluation metric $\varpi\in [0,1]$ as the ratio between the sum-rate w.r.t. a specific $\delta_0$ and the sum-rate without any angle mismatch. That is, the smaller the value of $\varpi$, the severer the performance degradation due to the angle mismatch. We see that the sum-rate degradation due to the angle mismatch problem indeed can be mitigated when $K$ decreases. For example, at $\delta_0=1$ we can achieve about $62\%$ of the sum-rate of the perfect spatial information case for $K=2$, while this percentage is  reduced to about  $23\%$ for $K=8$.
}
\begin{figure}[!t]
	%
	\centering
	\includegraphics[width=3.1 in]{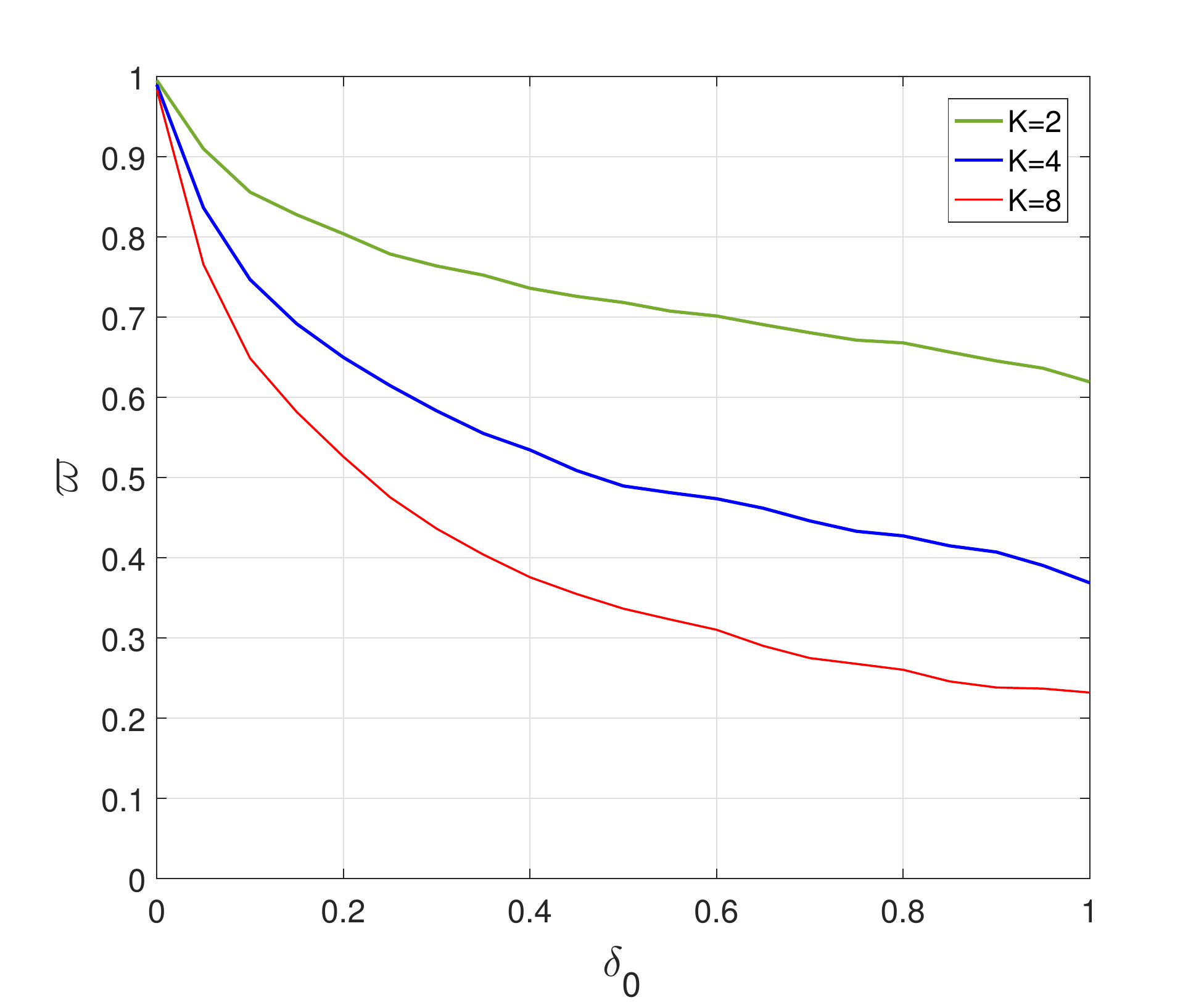}
	\caption{Sum-rate ratio $\varpi$ versus angle mismatch level for the proposed algorithm at $P=40$ dB and $N=64$. FS-based beam selection is adopted and $\Gamma$ is set to $1$.}
	\label{fig_rev}
\end{figure}

\begin{figure}[!t]
	\begin{minipage}[t]{0.5\linewidth}
		\centering
		\includegraphics[width=3.5 in]{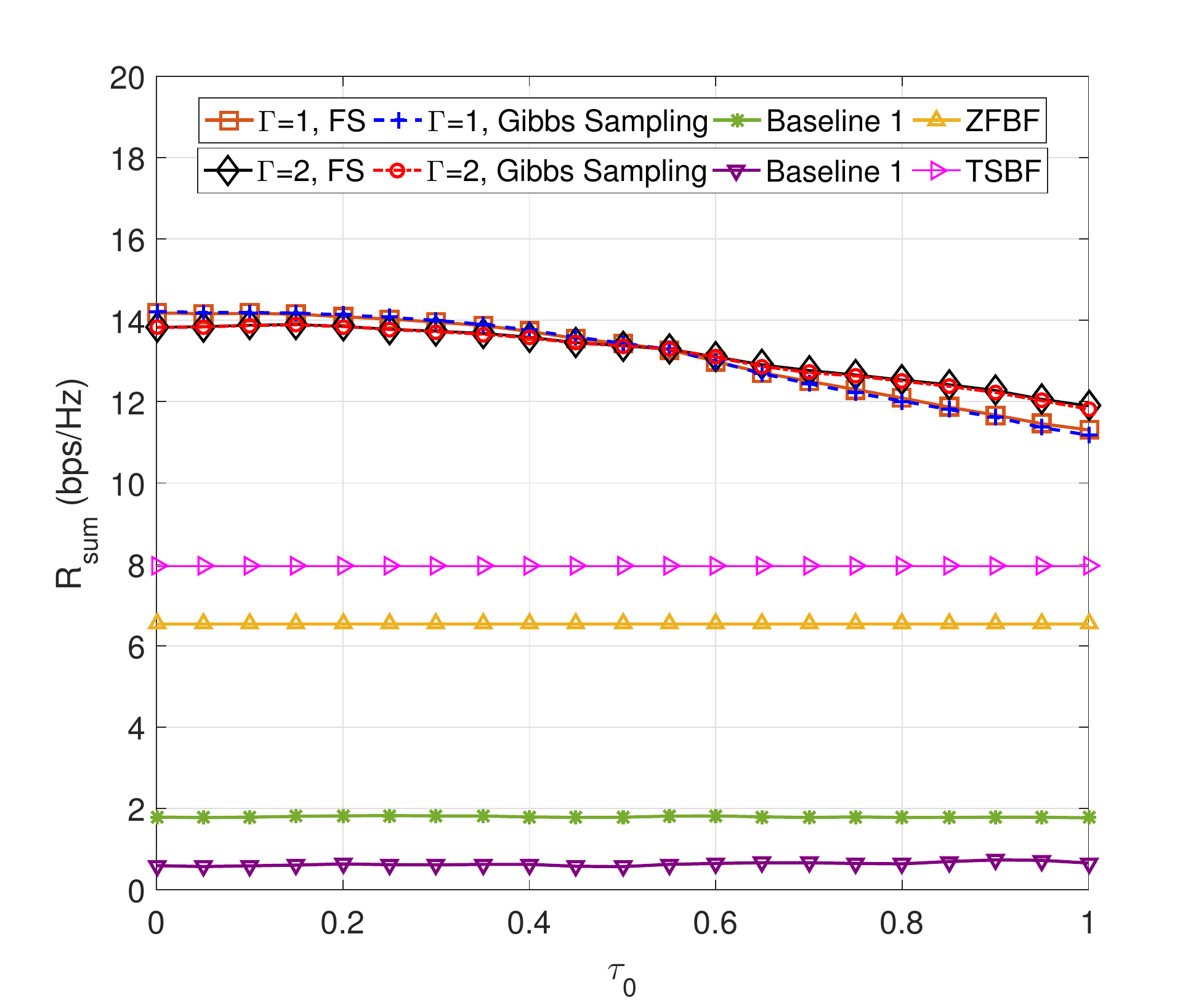}
		\caption{Performance for different PAS mismatch levels \protect\\at $P=20$ dB and $K=8$.}
		\label{fig_tau}
	\end{minipage}%
	\begin{minipage}[t]{0.5\linewidth}
		\centering
		\includegraphics[width=3.5 in]{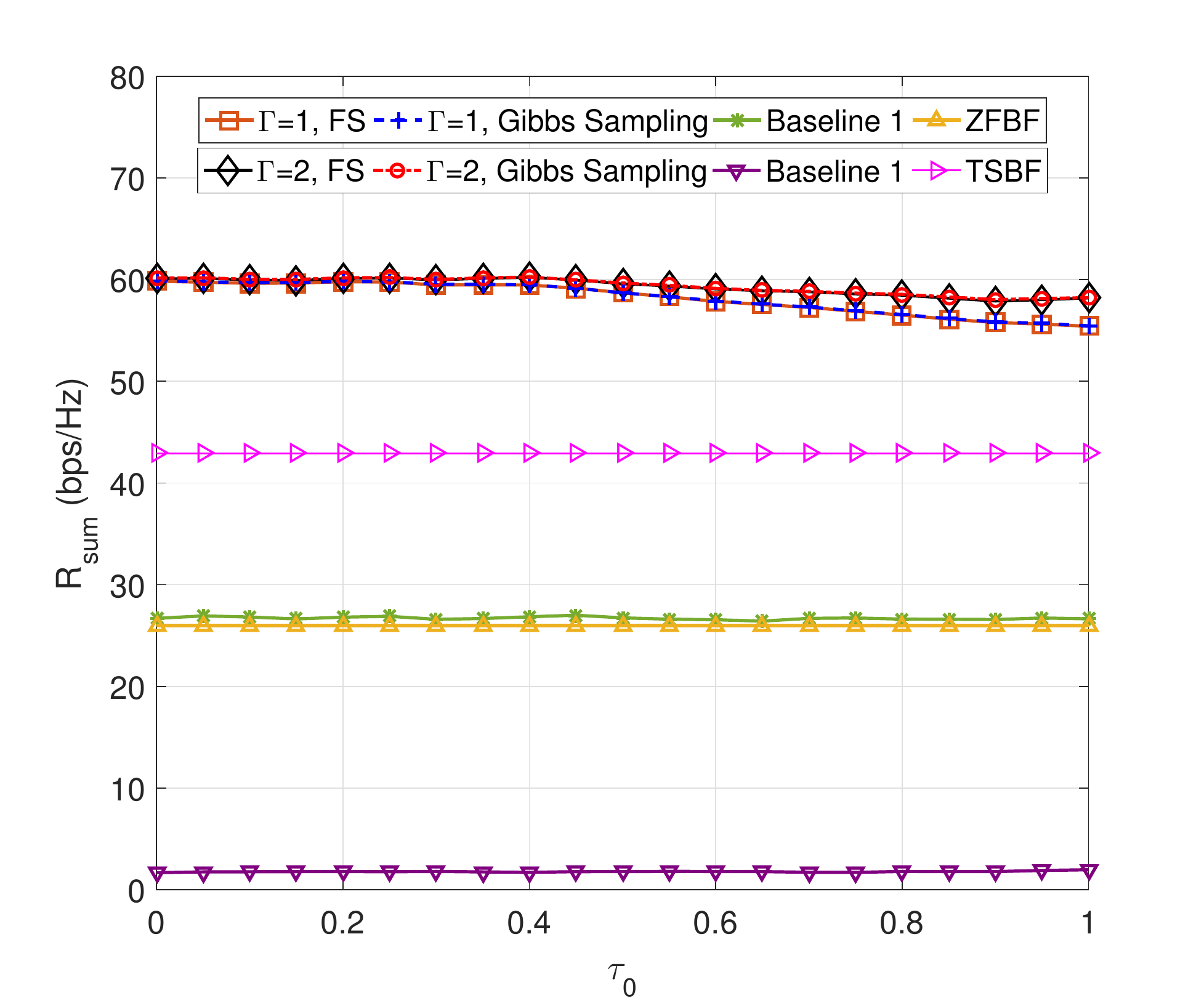}
		\caption{Performance for different PAS mismatch levels at $P=40$ dB and $K=8$.}
		\label{fig_tau2}
	\end{minipage}
\end{figure}

We now consider the impact of the PAS mismatch. Specifically, we set
\begin{align}\label{temp67}{\sigma^2_{k,l}\!=\!}
\begin{cases}
(1-\tau_0)\varsigma^2_{k,p}+\tau_0 \tilde \varsigma^2_{k,p}, &\text{if } \vartheta_l\!= \!\theta_{k,p} \text{ for some } \theta_{k,p}\! \in\! \thetav, \\
0, &\text{otherwise,}
\end{cases}
\end{align}
where $\tilde \varsigma^2_{k,p}$ represents the error in the PAS estimation and is drawn randomly following the same distribution as $\varsigma^2_{k,p}$, and the mismatch level $\tau_0$ controls the amount of the uncertain part in each $\sigma^2_{k,l}$. Figs. \ref{fig_tau} and \ref{fig_tau2} shows that the performance is robust w.r.t. a varying $\tau_0$, for that  $\sigma_{k,l}$ only affects the beam selection scheme and is independent of the calculation of the beamforming vector $\vv_k$ in \eqref{temp66}. 

\subsection{Simulations Under Propagation Geometry with Scattering Clusters}
\begin{figure}[!t]
	%
	\centering
	\includegraphics[width=3.1 in]{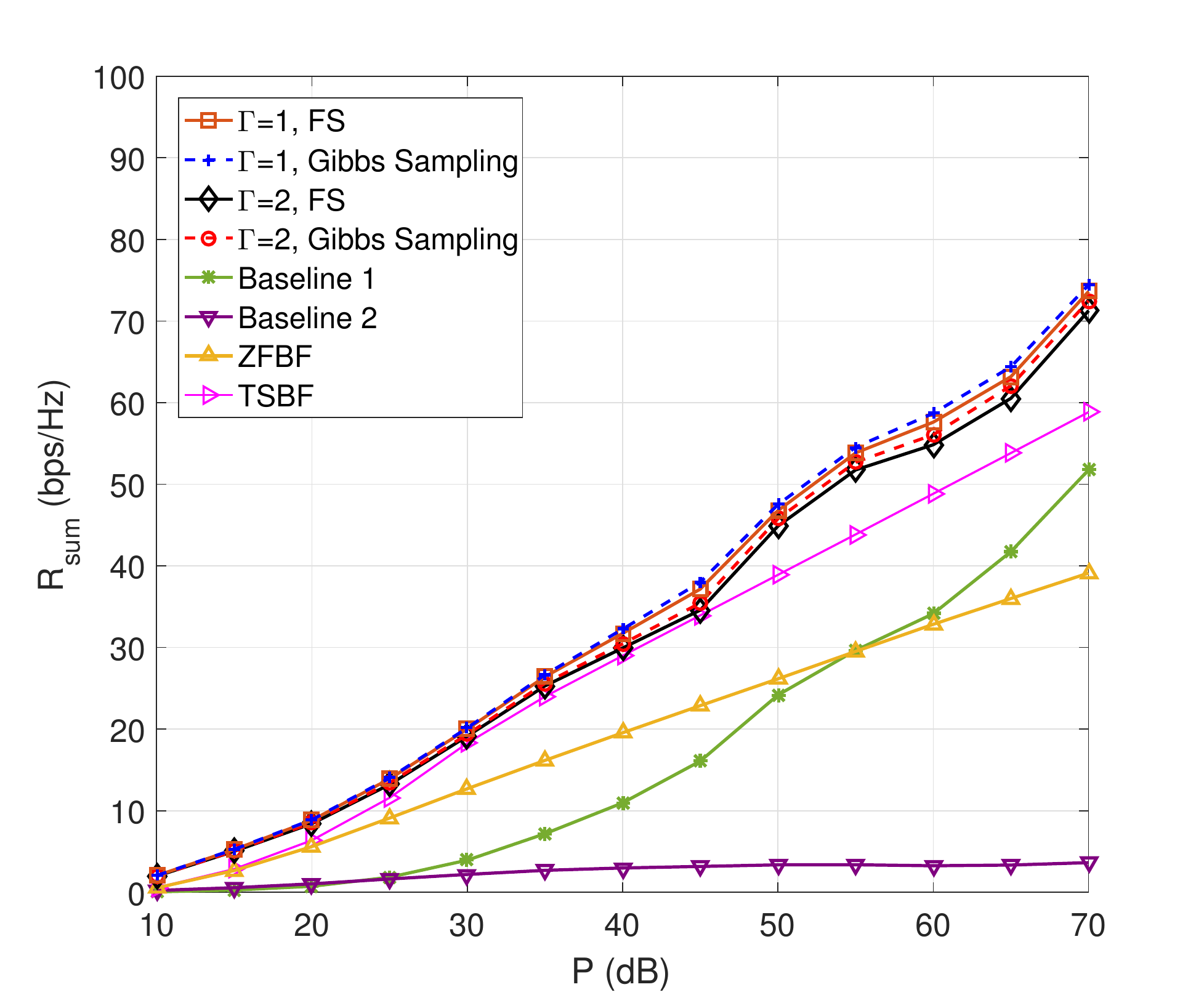}
	\caption{Sum-rate versus transmission power where the propagation environment is formed by $3$ scattering clusters.}
	\label{fig_rev2}
\end{figure}
{\blue
In this subsection, we consider the simulation geometry where the propagation environment is formed by the multiple multipath components (MPCs), each corresponding to a scattering cluster with certain angular spread. We assume $3$ MPC clusters with each center angle (parameterized by $\sin\theta$ instead of $\theta$) randomly drawn from $[-1,1]$. The size of each cluster is set to $0.4$. We set $K=5$, $N=60$, $L=30$, and $T=100$. The true angles for user $k$, denoted by $\thetav_k$, is randomly chosen from two of the three clusters (\ie any two users share at least one common cluster). The number of paths $P(k)$ for user $k$ is randomly selected from $[2,13]$. We assume perfect spatial information same as in Section VII-D. Fig. \ref{fig_rev2} plots the sum-rate performance of the proposed methods versus the transmission power $P$.
We see that the proposed algorithms have less performance improvements over the baselines compared with the result in Fig. \ref{P_R_sum}. This is because the considered geometry leads to large beam overlaps among the users, resulting in severe IUI. 
}
\section{Conclusions}
\label{conclusions}
In this paper, we studied the downlink beamforming design with no instantaneous CSIT in the FDD massive MIMO system. We presented a unified framework for spatial information acquisition by utilizing the channel parameters estimated in the uplink stage. Based on this, we proposed the BS-SBF design with the associated BTBC scheme. Moreover,  two beam selection methods were developed to maximize the approximate sum-rate, based on the FS search and the Gibbs sampling. Besides, we derived the sum-rate expression and its asymptotic limits. We also provided guidance on the system design by analyzing the effect of hyperparameters $\Gamma$ and $K$ on the sum-rate performance. Finally, numerical results demonstrate the significant performance improvement of our proposed algorithms compared to the existing schemes.
\appendices
\section{Proof of Proposition \ref{pro1}}
\label{appa}
Under the SINR expression \eqref{SINR_Gamma2}, the achievable rate of user $k$ is given by
\begin{align}
R_k&=\E\left[\log_2 \left( 1+\frac{\gamma}{\Upsilon\Gamma}\sum_{l \in \Omega_{k}} \sigma_{k,l}^2\abs{ s_{k,l}}^2\right) \right]-\E\left[\log_2 \left( 1+\frac{\gamma}{\Upsilon\Gamma}\sum_{l^\prime \in \Omega_{k}\setminus\mathcal{G}_k} \sigma_{k,{l^\prime}}^2\abs{ s_{k{,l^\prime}}}^2\right) \right].
\end{align}
Define $\varrho_k\triangleq\sum_{l \in \Omega_{k}} \sigma_{k,l}^2\abs{ s_{k,l}}^2 $. If $\Omega_{k}\neq \emptyset$, the probability density function (pdf) of $\varrho_k$ is given by
\cite[eq. (6)]{Chi-square-pdf}
\begin{align}
\label{temp54}
p_{\varrho_k}(\varrho)=\left(\prod_{j=1}^J \frac{1}{\sigma^{2r_j}_{k,j}} \right) \sum_{j=1}^J \sum_{l=1}^{r_j}\frac{(-1)^{2r_j-l-1}}{(r_j-l)!}\varrho^{r_j-l}f_{\rv,j,l}\exp\left(-\varrho/\sigma^2_{k,j} \right) ,\varrho\geq 0.
\end{align}
Besides, we have the following integral formula (cf. \cite[eq. (72)]{SBF_1.5}):
\begin{align}
\label{temp55}
\int_{0}^\infty\! \log_2(1+\frac{\gamma}{\Upsilon\Gamma}\varrho)\varrho^{r_j-l}\exp\left(-\varrho/\sigma^2_{k,j} \right)d\varrho\!=\!\frac{(r_j-l)!}{\ln 2}(\sigma^2_{k,j})^{r_j-l+1}\exp\left(\frac{\Upsilon\Gamma}{\gamma\sigma^2_{k,j}} \right) \sum_{t=1}^{r_j-l+1}\!\!\! E_{t}\left( \frac{\Upsilon\Gamma}{\gamma\sigma^2_{k,j}}\right).
\end{align}
Combining \eqref{temp54} and \eqref{temp55} gives the expression for $\E\left[\log_2 \left( 1+\frac{\gamma}{\Upsilon\Gamma}\sum_{l \in \Omega_{k}} \sigma_{k,l}^2\abs{ s_{k,l}}^2\right) \right]$ that corresponds to the first term on the r.h.s. of \eqref{Rk_us_multi}. Similar arguments apply to $\E[\log_2 ( 1+\frac{\gamma}{\Upsilon\Gamma}\sum_{l^\prime \in \Omega_{k}\setminus\mathcal{G}_k}$ $\sigma_{k,{l^\prime}}^2\abs{ s_{k{,l^\prime}}}^2)]$, yielding the second term in the r.h.s. of \eqref{Rk_us_multi}.

\section{\label{appa0}}
{\blue
At first, we consider a situation where the available beam sets (\ie $\{\mathcal{S}_k\}$ defined in \eqref{sk}) are pairwise disjoint. In this case, we must have $\mathcal{G}_k\subset \mathcal{S}_k, \forall k$ to maximize the sum-rate in \eqref{beamselection}. As a consequence, there is no IUI and $\text{SINR}_k$ in \eqref{SINR_Gamma2} is simplified as $\text{SINR}_k=\frac{\gamma}{\Upsilon\Gamma}\sum_{l\in \mathcal{G}_k}\sigma^2_{k,l}\abs{s_{k,l}}^2$. Therefore, the achievable rate of user $k$ is only related to  $\mathcal{G}_k$ and is independent of the beam choices of the other users. In this special case, we can simplify \eqref{beamselection} as
\begin{align}
\label{beamselection2}
(\mathcal{G}_1,\cdots,\mathcal{G}_K)\! =\!\argmax_{\substack{\mathcal{G}_i \bigcap \mathcal{G}_j =\emptyset, \forall i,j \\\abs{\mathcal{G}_k}=\Gamma,\forall k}} \sum_{k=1}^K R_{k}(\mathcal{G}_k).
\end{align}
As a result, \eqref{beamselection2} is equivalent to the (weighted) $\Gamma$-set packing problem: Given a set $[L]$, we aim to find $K$ subsets $\mathcal{G}_1,\cdots,\mathcal{G}_K$ to maximize the sum of weights assigned to the subsets (\ie $\sum_{k=1}^K R_{k}(\mathcal{G}_k)$), such that all the subsets are of the same size $\Gamma $ and are pairwise disjoint. It is shown in \cite{SetPacking} that the $\Gamma$-set packing problem is NP-hard if $\Gamma\geq 3$ and is equivalent to finding the maximal matching over a graph if $\Gamma=2$.  Since the $\Gamma$-set packing problem can be converted to a special case of \eqref{beamselection}, we conclude that solving \eqref{beamselection} is \emph{at least} as hard as solving the $\Gamma$-set packing problem. That is, the beam selection problem in \eqref{beamselection} is NP-hard if $\Gamma \geq3$, and is \emph{at least} as hard as solving the maximal  matching problem.
}
\section{Proof of Proposition \ref{pro2}}
\label{appb}
Utilizing the inequalities $\frac{x}{x+1}\leq \ln(1+x)\leq x$, for $0<\varrho<\infty$ we obtain
\begin{align}
\frac{\varrho}{\ln 2(1+\gamma\varrho)}\leq\frac{\log_2(1+\gamma \varrho)}{\gamma}\leq \frac{\varrho}{\ln2}.
\end{align}
This implies
\begin{align}
\label{temp56}
\lim_{\gamma\to 0}\frac{\log_2(1+\gamma \varrho)}{\gamma}= \frac{\varrho}{\ln2}.
\end{align}
From \eqref{SINR_Gamma2} and \eqref{temp56}, we have
\begin{align}
\lim_{\gamma\to 0}\frac{R_k}{\gamma}&\!=\!\lim_{\gamma\to 0}\frac{\E\left[\log_2 \left( 1+\frac{\gamma}{\Upsilon\Gamma}\sum_{l \in \Omega_{k}} \sigma_{k,l}^2\abs{ s_{k,l}}^2\right) \right]}{\gamma}\!-\!\lim_{\gamma\to 0}\frac{\E\left[\log_2 \left( 1+\frac{\gamma}{\Upsilon\Gamma}\sum_{l^\prime \in \Omega_{k}\setminus\mathcal{G}_k} \sigma_{k,{l^\prime}}^2\abs{ s_{k{,l^\prime}}}^2\right) \right]}{\gamma}\nonumber\\
&\!=\!\frac{\E\left[\sum_{l\in \mathcal{G}_k}\sigma^2_{k,l}\abs{s_{k,l}}^2\right]}{\Upsilon\Gamma \ln 2}=\frac{\sum_{l\in \mathcal{G}_k}\sigma^2_{k,l}}{\Upsilon\Gamma \ln 2}.
\end{align}

\section{Proof of Proposition \ref{pro3}}
\label{appc}
From \cite[eqs. (37), (38)]{SBF_2.5}, we have the following limit:
\begin{align}
\label{temp60}
\frac{1}{\ln 2}E_1\left( \frac{\Upsilon\Gamma}{\gamma \sigma^2_{k,l}}\right) \overset{\gamma \to \infty}\longrightarrow \log_2 (\gamma)+\log_2\left(\frac{\sigma^2_{k,l}}{\Upsilon\Gamma}\right)-\frac{\tilde{\gamma}}{\ln 2},
\end{align}
where $\tilde \gamma\approx 0.577$ is the Euler-Mascheroni constant.

Combining \eqref{temp60} and \eqref{Rk_us}, we obtain \eqref{highsnr}, where
\begin{subequations}
	\begin{align}
	C_1&=\sum_{k=1}^K\left(  \sum_{l \in \Omega_k}
	\left( \frac{1}{\prod_{\substack{j \neq l\\j \in \Omega_k}} \left( 1-\frac{\sigma^2_{k,j}}{\sigma^2_{k,l}}\right) } \right)\!-\! \sum_{l^\prime \in \Omega_k\setminus \mathcal{G}_k}
	\left( \frac{1}{\prod_{\substack{j \neq l^\prime\\j \in \Omega_k\setminus\mathcal{G}_k}} \left( 1-\frac{\sigma^2_{k,j}}{\sigma^2_{k,{l^\prime}}}\right) } \right)\right),\\
	C_2&=\sum_{k=1}^K\left(  \sum_{l \in \Omega_k}
\left( \frac{\log_2\left(\frac{\sigma^2_{k,l}}{\Upsilon\Gamma}\right)\!-\!\frac{\tilde{\gamma}}{\ln 2}}{\prod_{\substack{j \neq l\\j \in \Omega_k}} \left( 1-\frac{\sigma^2_{k,j}}{\sigma^2_{k,l}}\right) } \right)\!-\! \sum_{l^\prime \in \Omega_k\setminus \mathcal{G}_k}
\left( \frac{\log_2\left(\frac{\sigma^2_{k,l^\prime}}{\Upsilon\Gamma}\right)\!-\!\frac{\tilde{\gamma}}{\ln 2}}{\prod_{\substack{j \neq l^\prime\\j \in \Omega_k\setminus \mathcal{G}_k}} \left( 1-\frac{\sigma^2_{k,j}}{\sigma^2_{k,{l^\prime}}}\right) } \right)\right).
	\end{align}
\end{subequations}

\ifCLASSOPTIONcaptionsoff
\newpage
\fi
\ifhavebib
{
	\bibliographystyle{IEEEtran}

}
\else{
}
\fi
\end{document}